\documentclass[reqno,12pt,a4paper]{amsart}
\usepackage{bbm}
\usepackage{amsmath} % loades automatisk i klassen 'amsart'.

\usepackage{amstext}
\usepackage[danish]{babel}
\usepackage[latin1]{inputenc}
\usepackage{amsthm}

\title{Membranes and Matrix Models}
\author{Jens Hoppe}
\newcommand{\unity}{{\setlength{\unitlength}{1em}
                    \begin{picture}(0.75,1)
                    \put(0,0){$1$}
                    \put(0.38,0){\line(0,1){0.65}}
                    \end{picture}
                  }}

\newcommand{\R}{{\mathbb{R}}}
\newcommand{\N}{{\mathbb{N}}}
\newcommand{\Q}{{\mathbb{Q}}}
\newcommand{\C}{{\mathbb{C}}}

\begin{document}
    
\maketitle

%\begin{center}
%\subsection*{MEMBRANES AND MATRIX MODELS}
%\end{center}

%\bigskip

%\begin{center}
%    {\sc Jens Hoppe}
%\end{center}

%\bigskip

\begin{quote}
{ \small      Royal Institute of Technology, S-10044 Stockholm, and \\      
IHES, 35, route 
de Chartres, F-91440 Bures-sur-Yvette. \\
     E--mail: {\tt hoppe@math.kth.se}}
\end{quote}\bigskip\bigskip

For many years, the derivation of
\[
H = - \triangle - Tr \sum_{i<j} [X_{i}, X_{j}]^2
\]
(involving finitely many antihermitean traceless $N \times N$ matrices 
$X_{i} = \displaystyle\sum_{a=1}^{N^2-1} x_{ia} T_{a},\ i=1,\dots, 
d)$\footnote{i.e. $H = - \frac{\partial}{\partial x_{ia}}\ 
\frac{\partial}{\partial x_{ia}} + \frac{1}{2}\ f^{(N)}_{abc} 
f^{(N)}_{ab'c'} x_{ib}x_{jc}x_{ib'}x_{jc'}$, for each $N$ and $d$, 
being an ordinary Schr\"odinger--operator on ${\R}^{d(N^2-1)}$, with a 
non--negative quartic potential, given in terms of totally 
antisymmetric $su(N)$--structure--constants $f^{(N)}_{abc} = - Tr 
T_{a} [T_{b}, T_{c}]$.} as
describing a quantized discrete analogue of relativistically 
invariant surface dynamics\footnote{as well as reduced Yang Mills theory} 
in ${\R}^{d+1}$ was available only in rather inconvenient 
forms (handwritten [1], too condensed [2], typeset\footnote{with
unfortunately many typos} by a local journal [3]). With the 
supersymmetric analogue of $H$,
\[
H_{\text{Susy}} = H \cdot \unity + i f_{abc} x_{jc} 
\gamma_{\alpha\beta}^{j} \theta_{\alpha a} \theta_{\beta b}\ ,
\]
$\{\theta_{\alpha a}, \theta_{\beta b}\} = 
\delta_{\alpha\beta}\delta_{ab},\ (\gamma^j \gamma^k + \gamma^k 
\gamma^j)_{\alpha\beta}=2\delta_{\alpha\beta}\delta^{jk}$, cf [4, 5], 
having become relevant for yet another reason [6], - and for the 
participants of a summer school on Schr\"odinger--operators, it seemed 
useful to give a detailed account of $H$ (Section II) 
while adding various introductory remarks about surface motions, their relation 
with hydrodynamics [7], as well as diffeomorphism -- and
relativistic invariance (Section I).

Section III concerns a rather particular, though quite important, 
question: does $H_{\text{Susy}}$ (whose spectrum is known to cover the
whole positive axis, ${\R}^+$ [8, 29]) admit a zero energy bound state, or
not? Some work on this is summarized.

Section IV (following [9], [10], [11]) 
discusses the space of solutions of the differential equations
\[
\dot{X}_{a} = \epsilon_{abc} X_{b} X_{c} - 2 X_{a}
\]
for $3$ traceless anti--hermiteam $N \times N$ matrices $X_{a}(t),\ t 
\in (-\infty , + \infty)$, interpolating between different 
representations of $su (2)$.

Some excercises have been added, and a remark/conjecture concerning 
5-commutators.
\\[1em]

I am very grateful to M. Ring and J.P. Solovej, as well as MaPhySto 
and IHES,
for making possible this written version of 4 lectures presented at 
Sandbjerg Castle during the summer school on `Quantum Field Theory -- 
from a Hamiltonian Point of View', August 2--9, 2000. \bigskip

\newpage

\begin{center}
 {\bf I}
\end{center}

Let me start with two relatively simple surface motions,
{\def\theequation{\arabic{equation}}
\setcounter{equation}{0}
\begin{eqnarray}
\dot{\stackrel{\rightharpoonup}{x}} & = & 
\stackrel{\rightharpoonup}{n}\ , \\
\dot{\stackrel{\rightharpoonup}{x}} & = & \sqrt{g}\ 
\stackrel{\rightharpoonup}{n} ( = \partial_{1} 
\stackrel{\rightharpoonup}{x} \times \partial_{2} 
\stackrel{\rightharpoonup}{x})\ .
\end{eqnarray}}
In both cases the surface is, for each time $t$, described in a 
parametric way (by giving $\stackrel{\rightharpoonup}{x} (t, 
\varphi^1, \varphi^2))$, i.e. by viewing the surface $\sum_{t} 
\subset {\R}^3$ as the image of a (timedependent) map from some fixed 
manifold $\sum_{(2)}$ (specifying the topology; e.g. that of $S^2$) 
into ${\R}^3$. As long as $\sum_{t}$ is non--degenerate, every vector 
can be decomposed into it's components parallel to $\frac{\partial 
\stackrel{\rightharpoonup}{x}}{\partial 
\varphi^1}$ and $\frac{\partial 
\stackrel{\rightharpoonup}{x}}{\partial \varphi^2}$ (tangent 
to $\sum_{t}$) resp.
{\def\theequation{3}
\begin{equation}
    \stackrel{\rightharpoonup}{n} := 
    \frac{\partial_{1}\stackrel{\rightharpoonup}{x}\times 
    \partial_{2}\stackrel{\rightharpoonup}{x}}{|\partial_{1}
\stackrel{\rightharpoonup}{x}\times\partial_{2}\stackrel{\rightharpoonup}{
x}|}
\end{equation}}
(normal to $\sum_{t}$); $\partial_{r}\stackrel{\rightharpoonup}{x}$ 
stands for $\frac{\partial 
\stackrel{\rightharpoonup}{x}}{\partial \varphi^r}, r =1,2, 
\hspace{1ex} g$ denotes 
the determinant of the $2\times 2$ matrix 
$(g_{rs})=(\partial_{r}\stackrel{\rightharpoonup}{x} \cdot 
\partial_{s}\stackrel{\rightharpoonup}{x})$.

The evolution equation (1), though consisting of coupled 
non--linear PDE's for the $3$ unknown functions $x^{i}(t, \varphi^1, 
\varphi^2)$ is trivial, as $\stackrel{\rightharpoonup}{n}^2 = 1$ implies 
$\stackrel{\rightharpoonup}{n} \cdot 
\dot{\stackrel{\rightharpoonup}{n}}=0$ and 
$\stackrel{\rightharpoonup}{n} \cdot \partial_{r} 
\stackrel{\rightharpoonup}{n}=0$, hence
{\def\theequation{4}
\begin{equation}
    \ddot{\stackrel{\rightharpoonup}{x}} = 
    \dot{\stackrel{\rightharpoonup}{n}} = 0
\end{equation}} \vspace{-0.5cm}
\[
    \stackrel{\rightharpoonup}{x} (t, \varphi^1, \varphi^2) = 
    \stackrel{\rightharpoonup}{x} (0, \varphi^1, \varphi^2) + t\ 
    \stackrel{\rightharpoonup}{n} (0, \varphi^1, \varphi^2)\ ,
\]
as $\dot{\stackrel{\rightharpoonup}{n}}$ has zero component in the 
direction of $\stackrel{\rightharpoonup}{n}$, as well as parallel to 
$\sum_{t} \hspace{1ex} (\dot{\stackrel{\rightharpoonup}{n}} \partial_{r} 
\stackrel{\rightharpoonup}{x} = - \stackrel{\rightharpoonup}{n} 
\partial_{r} \dot{\stackrel{\rightharpoonup}{x}} = - 
\stackrel{\rightharpoonup}{n} \partial_{r}\stackrel{\rightharpoonup}{n})$.

What about (2)?

Despite the r.h.s. being polynomial in the first derivatives (so, from 
this point of view, being `simpler' than (1)) the (still non--linear) 
equation (2) needs quite different techniques to be `solved'. One 
possibility is to note that after interchanging dependent and 
independent variables, $t$ (as a function of 
$\stackrel{\rightharpoonup}{x}$!) turns out to satisfy a linear! (and, 
in fact, the simplest possible $2^{\text{nd}}$ order) differential 
equation,
{\def\theequation{5}
\begin{equation}
    \triangle t (\stackrel{\rightharpoonup}{x}) = 0\ ;
\end{equation}}
so the time at which the surface passes a point 
$\stackrel{\rightharpoonup}{x}$ in space is a harmonic function 
(resp. $\sum_{t}$ level sets of that harmonic function) [12, 13].

What if the normal velocity (= $\sqrt{g}$ in (2)) is generalized to 
be an arbitrary (non--linear) function of $\sqrt{g}$?\footnote{To 
avoid (hopefully not cause) confusion: strictly speaking, $\sqrt{g}$ 
always has to be divided by some reference--density $\rho$ in order 
to make the evolution equation (2), resp. (6), well defined (see [13] 
for a full account of this).}

As shown in [13], the evolution equations
{\def\theequation{6}
\begin{equation}
\dot{x}^{i} = \alpha (\sqrt{g})\ n^{i}\ ,
\end{equation}}
for arbitrary monotonic function $\alpha$, and the hypersurface 
motion taking place 
in any Riemannian manifold ${\mathcal{N}} (i = 1 \dots M + 1$ if $x^{i}
= 
x^{i} (t, \varphi^1, \dots , \varphi^{M})$, and $g = \det 
(\partial_{r}x^{i} \partial_{s} x^j \eta_{ij}(x),\ \eta_{ij}=$ metric 
on ${\mathcal{N}})$ can always be converted to a second--order equation 
for $t(x)$ (however, only in the case (2) this equation becomes 
linear).

Furthermore, despite their appearance as first order equations (in the 
derivatives of the $x^{i}$) they can be viewed as coming from a class 
of diffeomorphism invariant Hamiltonians [13] [14];
{\def\theequation{7}
\begin{equation}
    H [x^{i}, p_{j}] = \int d^{M} \varphi \sqrt{g} h (p / \sqrt{g})
\end{equation}}
$(p = \sqrt{p_{i}p_{j}\eta^{ij}(x)})$, restricted to
$C_{r} := p_{i} \partial_{r} x^{i},\hspace{1ex} r = 1, \dots , M,$
(the generators of diffeomorphisms, which are constants of the 
motion) set equal to zero!

Returning now to flat embedding space and $M=2$ ($2$--surfaces moving 
in ${\R}^3$) for simplicity, let me ask the following question: Does 
there exist a function $h$ (corresponding to the freedom of choosing 
$\alpha$ in (6)) such that the corresponding surface motions are 
relativistically invariant, i.e. for which $H$ can be complemented by 
$9$ other functionals of $\stackrel{\rightharpoonup}{x}$ and 
$\stackrel{\rightharpoonup}{p}$,
altogether generating the 
inhomogeneous Poincar\'{e}--group?

The answer is `Yes': for $h(u) = \sqrt{u^2+1}$, i.e.
{\def\theequation{8}
\begin{equation}
   H = \int_{\sum_{(2)}} d^2 \varphi 
   \sqrt{\stackrel{\rightharpoonup}{p}^2+g}
\end{equation}}
one not only has
{\def\theequation{9}
\begin{equation}
   L_{ij} :=  \int_{\sum_{(2)}} d^2 \varphi (x_{i} p_{j} - x_{j} 
   p_{i})\ , \hspace{3ex} \stackrel{\rightharpoonup}{p} :=
\int_{\sum_{(2)}} d^2 
   \varphi \stackrel{\rightharpoonup}{p}
\end{equation}}
(generating rotations, and spatial translations), but also the 
generators of `boosts',
{\def\theequation{10}
\begin{equation}
  K_{i} := \int_{\sum_{(2)}} d^2 \varphi x_{i} 
  \sqrt{\stackrel{\rightharpoonup}{p}^2+g}\ .
\end{equation}}
Restricting to $C_{r} := \stackrel{\rightharpoonup}{p} \cdot 
\partial_{r} \stackrel{\rightharpoonup}{x}=0$ then gives 
$\sqrt{\stackrel{\rightharpoonup}{p}^2+g} = \rho =$ 
time--independent density, and -- using this --
{\def\theequation{11}
\begin{equation}
 \dot{\stackrel{\rightharpoonup}{x}} = \pm \sqrt{1-g/\rho^2}\ 
 \stackrel{\rightharpoonup}{n}.
\end{equation}}
$2$--dimensional surfaces moving according to (11) not only 
correspond to `relativistically invariant' motions, but have the more 
specific property that they actually sweep out a $3$--dimensional 
manifold ${\mathcal{M}}$ in Minkowski--space, which has vanishing mean 
curvature (see e.g. [15], and below).

-- In any case, comparing (8), (9), (10) with the corresponding 
expressions for a (finite--dimensional) system of $N$ free (!) 
relativistic particles
{\def\theequation{\arabic{equation}}
\setcounter{equation}{11}
\begin{eqnarray}
H & = & \sum_{a=1}^{N} \sqrt{\stackrel{\rightharpoonup}{p}^2_{a}  + 
m^2_{a}}\ , \hspace{4ex} \stackrel{\rightharpoonup}{P} \hspace{1.5ex} =
\hspace{1ex} \sum_{a} 
\stackrel{\rightharpoonup}{p}_{a}\nonumber\\[0.5em]
\stackrel{\rightharpoonup}{L} & = & \sum_{a} 
(\stackrel{\rightharpoonup}{x}_{a} \times 
\stackrel{\rightharpoonup}{p}_{a})\ , \hspace{4ex} 
\stackrel{\rightharpoonup}{K} \hspace{1.5ex} = \hspace{1ex} \sum_{a} 
\stackrel{\rightharpoonup}{x}_{a}
\sqrt{\stackrel{\rightharpoonup}{p}^2_{a}  + 
m^2_{a}}\ ,
\end{eqnarray}}
one finds exact correspondance ($a \stackrel{\wedge}{=} \varphi^1, 
\varphi^2)$ when replacing the position--independent masses $m_{a},\ 
a = 1, \dots , N$, by the position--{\em dependent} density 
$\sqrt{g[\stackrel{\rightharpoonup}{x}(\varphi)]}$.
While it was proved decades ago [16] that, given certain 
physical requirements concerning the realization of the $10$ 
generators, any set deviating from (12), but still satisfying the 
same commutation relations, can, by a sequence of canonical 
transformations, be brought into the form (12) (so in particular, it 
is, for finite $N$ not! possible to allow for $x$--dependent masses 
$m_{a}$) it should be noted that the way (9)--(10) circumvents the 
just mentioned No--Interaction theorem for finitely many degrees of 
freedom, is quite interesting.

Another notable aspect is the possibility to abandon the 
parametric description at this point and describe the surface 
motion, in Hamiltonian form, purely in terms of differeomorphism 
invariant objects. Formally one gets, as a reduced phase--space, the 
space of shapes $\sum$ (as the configuration manifold) together with 
functions on $\sum$ (elements of the cotangentspace at $\sum$) as 
`momenta'. In this formulation, the generators of the 
Poincar\'{e}--group read [17]
{\def\theequation{\arabic{equation}}
\setcounter{equation}{12}
\begin{eqnarray}
\Bbb{H} \bigl[\Sigma, u\bigr] & = & \int_{\sum} \sqrt{u^2+1}\ ,
\hspace{12ex} 
\stackrel{\rightharpoonup}{\Bbb{P}} \hspace{1.5ex} = \hspace{1ex} 
\int_{\sum} \stackrel{\rightharpoonup}{n} u
\nonumber\\[0.5em]
\Bbb{L}_{ij} & = & \int_{\sum} (x_{i} n_{j} - x_{j} n_{i})\ u\ , 
\hspace{4ex} \Bbb{K}_{i} \hspace{1.5ex} = \hspace{1ex} 
\int_{\sum} x_{i} \sqrt{u^2+1}\ .
\end{eqnarray}}
Amazingly, the Poisson structure is the canonical one, meaning that the 
equations of motion are
{\def\theequation{14}
\begin{equation}
 \dot{\Sigma} = \frac{\delta \Bbb{H}}{\delta u} = 
 \frac{u}{\sqrt{u^2+1}}\ , \hspace{4ex} \dot{u} = - \frac{\delta 
 \Bbb{H}}{\delta\Sigma} = - 
 \sqrt{u^2+1}\ H
\end{equation}}
i.e. (as $\dot{\sum}$, the time--derivative of $\sum$, can be 
identified with the normal velocity, $v$)
{\def\theequation{15}
\begin{equation}
  \dot{v} = - (1-v^2)\ H\ ;
\end{equation}}
here $H$ denotes the mean 
curvature of $\sum$ (which in a parametric description equals 
$- g^{rs}\stackrel{\rightharpoonup}{n}\cdot \partial^2_{rs} 
\stackrel{\rightharpoonup}{x}$)\footnote{For the reader familiar with 
the fact that the first variation of the area--functional gives $H$, 
the second part of (14) may become intuitively clear, when first 
dropping the argument $\sqrt{u^2+1}$ in $\Bbb{H}$ (and (14)).}

Writing (15) as 
$\frac{\dot{v}}{(1-v^2)^{3/2}}+\frac{H}{(1-v^2)^{1/2}}=0$ one finds 
that the $3$--manifold ${\mathcal{M}}_{3}$ swept out in space--time by a
surface 
moving according to (15) will actually have {\em vanishing} mean 
$3$--curvature $H_{3}$ (the spatial part, proportional to $H$, is
cancelled 
by the curvature of the wordline of the point on $\sum$; the 
$4$--dimensional hypersurface--normal is, up to orientation, 
$n^\mu = ( \frac{v}{\sqrt{1-v^2}}, 
\frac{\stackrel{\rightharpoonup}{n}}{\sqrt{1-v^2}}))$.

Conventionally, the problem of finding such $3$--manifolds 
${\mathcal{M}}_{3}$ is formulated as considering (in a Diff 
${\mathcal{M}}_{3}$--invariant, parametric, way)
{\def\theequation{\arabic{equation}}
\setcounter{equation}{15}
\begin{eqnarray}
S [x^\mu] & = & \text{Vol}\; ({\mathcal{M}}_{3}) = \int d^3 \varphi 
\sqrt{G} \nonumber \\[0.5em]
G & = & \det\; \bigl( \frac{\partial x^\mu}{\partial \varphi^\alpha} 
\frac{\partial x^v}{\partial \varphi^\beta} \eta_{\mu 
v}\bigr)_{\alpha,\beta=0,1,2} \\[0.5em]
(\eta_{\mu v}) & = & \text{diag}\; (1, -1, -1, -1)\ , \nonumber
\end{eqnarray}}
a functional of the embedding functions
$x^\mu (\varphi^0, \varphi^1, \varphi^2)\
\text{(describing ${\mathcal{M}}_{3}$),}$ whose first variation
($\stackrel{!}{=} 0$) gives the equations of motions
{\def\theequation{17}
\begin{equation}
  \frac{1}{\sqrt{G}}\ \partial_{\alpha}\ \sqrt{G}\ G^{\alpha\beta} 
  \partial_{\beta} x^\mu = 0\ .
\end{equation}}
For a direct derivation of (8), resp. (11), from (16), resp. (17), see 
e.g. [15].

Alternatively, (16) could have been motivated as follows: in order to 
describe a relativistically invariant motion of an extended object (a 
surface, in this case), the only chance is to consider the manifold 
$\mathcal{M}$ swept out in space--time, and demand some 
extremality--property of $\mathcal{M}$; considering the 
Volume--functional is singled out by the fact that, 
depending only on the first derivatives of $x^\mu$, the equations of 
motion, (17), are then of second order (other 
diffeomorphism--invariant functionals like the total curvature of 
${\mathcal{M}},\dots,$ would involve second or higher order derivatives
of the 
$x^\mu$, hence leading to higher order equations of motion).

Perhaps a third connecting path is worth mentioning: a Lagrangian 
formulation of (13) would give
{\def\theequation{\arabic{equation}}
\setcounter{equation}{17}
\begin{eqnarray}
S & = & \int dt \bigl( \int_{\sum} vu - \int_{\sum} 
\sqrt{u^2+1}\bigr) \nonumber \\[0.5em]
& = & - \int dt \int_{\sum} \bigl( \frac{1}{\sqrt{u^2+1}} = 
\sqrt{1-v^2} = \sqrt{1-\dot{\Sigma}^2}\bigr) \\[0.5em]
& = & S \bigl[\Sigma, \dot{\Sigma}\bigr]\ . \nonumber
\end{eqnarray}}
Now, describing $\sum = \sum (t)$ as the set of points $(t, 
\stackrel{\rightharpoonup}{x})$, where some functions $m(t, 
\stackrel{\rightharpoonup}{x})=m(x^\mu)$ vanishes, (18) becomes
{\def\theequation{\arabic{equation}}
\setcounter{equation}{18}
\begin{eqnarray}
S & = & - \int d^4 x (\delta (m(t, \stackrel{\rightharpoonup}{x}))\ 
|\stackrel{\rightharpoonup}{\nabla} m|)\ 
\sqrt{1-\frac{\dot{m}^2}{|\nabla m|^2}}\nonumber \\[0.5em]
& = & - \int d^4 x \delta (m) \sqrt{- \partial_{\mu} m \partial^\mu m}
\hspace{1ex} 
= \hspace{1ex} S[m]\ ,
\end{eqnarray}}
with equations of motion $\delta (m) \partial_{\mu} \bigl( 
\frac{\partial^\mu m}{\sqrt{-\partial_{\rho} m \partial^\rho m}}\bigr) = 
0$,
respectively
{\def\theequation{20}
\begin{equation}
  (\eta^{\mu\nu} \eta^{\rho \lambda} - \eta^{\mu \rho} 
  \eta^{\nu\lambda}) \partial_{\mu}m \partial_{\nu}m \partial^2_{\rho 
  \lambda}m = 0\ ,
\end{equation}}
together with $m(t, \stackrel{\rightharpoonup}{x}) \stackrel{!}{=} 0$;
meaning 
that first solving (20) and then solving for ${\mathcal{M}} := \{(t, 
\stackrel{\rightharpoonup}{x})\ |m(t, \stackrel{\rightharpoonup}{x}) 
=0\}$ will yield a $3$--manifold with everywhere vanishing mean 
curvature. 
To understand why the factors of
$|\stackrel{\rightharpoonup}{\nabla}_{m}|$ 
have to enter in $(\delta (m) 
|\stackrel{\rightharpoonup}{\nabla}_{m}|)$ and $(v^2 = 
\frac{\dot{m}^2}{|\nabla_{m}|^2})$ 
precisely as indicated, making both expressions, hence (20), invariant 
under $m \to F(m)$, one should note that if $m=0$ describes
$\mathcal{M}$, 
functions of $m$, like $F(m) = m^3$ would yield the same $\mathcal{M}$. 
Furthermore note that $m(t, \stackrel{\rightharpoonup}{x}) =$ const. 
$(\not= 0)$, with $m$ solving (20) also yields an 
extremal $3$--manifold.
Locally, Minkowski--space is therefore foliated into $3$ manifolds, 
each with vanishing mean curvature, with $m$ being the parameter 
orthogonal to the extremal $3$--manifolds. In physics, the direct 
correspondance of (20) and (17) was established by 
Sugamoto [18] and in [19] (see also [20], in particular concerning the 
factor $\delta (m)$).

{\bf An Example:} Making the Ansatz $m(t, 
\stackrel{\rightharpoonup}{x}) = \displaystyle\prod_{\mu=0}^{3} 
m_{\mu} (x^\mu)$, resp. $\widetilde{m} = F(m) = \ln m =
\displaystyle\sum_{0}^3 
\ln m_{\mu} = \displaystyle\sum_{0}^3 \widetilde{m}_{\mu}(x^\mu)$, one
can derive 
[15] that
{\def\theequation{21}
\begin{equation}
 \widehat{{\mathcal{M}}} := \{ (t, \stackrel{\rightharpoonup}{x})\ |\ 
 {\mathcal{P}}(x) 
 {\mathcal{P}}(y) {\mathcal{P}}(t) = {\mathcal{P}}(z)\}\ ,
\end{equation}}
where ${\mathcal{P}}(u)$ is an elliptic Weierstrass--function satisfying 
${\mathcal{P}}^2=4{\mathcal{P}}({\mathcal{P}}^2-1)$, defines a periodic 
$\mathcal{M}$ with vanishing mean 
curvature. Viewed as time evolutions of $2$--surfaces, (21) 
corresponds to (at $t=0, \pm\omega, \pm 2\omega,\dots$) infinitely 
many parallel planes, at distance $\omega$ apart from each other, 
breaking up into little squares of size $\omega \times \omega$ that 
grow one scalactite and one scalagmite in the direction 
perpendicular to the original planes meeting eventually, at times 
$t=\pm \frac{\omega}{2}, \pm \frac{3\omega}{2}, \dots$ at the center 
of each `box' of size $\omega \times \omega\times\omega$. \\

{\bf Note} [19]: \\
Just choosing $\varphi^\alpha = x^\alpha$ in (16), leaving $x^3 = z 
(x^0, x^1, x^2)$ to be determined, yields the same equation than 
writing \\ $m=m(x^0, x^1, x^2, z(x^0, x^1, x^2))$ in (20); choosing 
$\varphi^0=\frac{x^0+x^3}{2}=\tau,\ \varphi^r = x^r$ in (16), resp. 
writing $(\tau + \frac{1}{2}\ p (\tau, x^1, x^2), x^1, x^2, \tau - 
\frac{p}{2})$ in (20) both yields a second order equation for $p$, which 
in first order from (defining $q = (2\dot{p} + (\nabla 
p)^2)^{-\frac{1}{2}})$ becomes
{\def\theequation{\arabic{equation}}
\setcounter{equation}{21}
\begin{eqnarray}
\dot{q} + \stackrel{\rightharpoonup}{\nabla} (q 
\stackrel{\rightharpoonup}{\nabla} p) & = & 0 \\[0.5em]
\dot{p} & = & \frac{1}{2}\ \bigl(\frac{1}{q^2} - 
(\stackrel{\rightharpoonup}{\nabla} p)^2\bigr)\ , \nonumber
\end{eqnarray}}
the continuity, and Euler--equation for an inviscid, isentropic 
irrotational gas, with density $q$, velocity potential $p$, and 
equation of state (K\'{a}rm\'{a}n--Tsien gas) Pressure $= 
\frac{-1}{\text{Densitiy}}$.

This relation between relativistic membrane motions and
$2+1$--dimensional hydrodynamics, (22), was first derived in [7], and 
extended to a supersymmetric membrane/fluid in [21]. \\[1em]

While it is quite interesting to have all these different 
formulations (revealing different aspects of one and the same problem, 
-- hopefully leading, one day, to some hidden integrability structure) 
none of them could, up to now, be quantized. The best route to 
quantisation, known so far, is the one found $20$ years ago (cp. [1], 
resp. [22]) to be discussed after the following

\newpage

\noindent {\bf Excercise 1)}
\begin{enumerate}
   \item[ \ ] Choose, in (17), $\varphi^0 = \frac{x^0+x^3}{2} =:
\tau$, 
   and assume $(\varphi^1, \varphi^2)$ to be chosen such that
   
   $G_{or} = \partial_{r} \zeta - \dot{\stackrel{\rightharpoonup}{x}} 
   \partial_{r} \stackrel{\rightharpoonup}{x} \equiv 0$, where 
   $\stackrel{\rightharpoonup}{x} = (x^1, x^2),\ \cdot = 
   \frac{\partial}{\partial\tau}$ and $x^0-x^3 =: \zeta (\tau, 
   \varphi^1, \varphi^2)$.
   
   Verify that $\sqrt{G} = \sqrt{g}\ \sqrt{2 \dot{\zeta} - 
   \dot{\stackrel{\rightharpoonup}{x}}^2}$ and show that, as a 
   consequence of (17), i.e. $\triangle x^\mu \stackrel{!}{=} 0$,
  \begin{enumerate}
      \item[a)] $\rho :=
\frac{\sqrt{g}}{\sqrt{2\dot{\zeta}-\dot{\stackrel{\rightharpoonup}{x}}^2}}
$ 
      is independent of $\tau$ (hint: $\triangle (x^0+x^3)=0)$
      \item[b)] $\ddot{\stackrel{\rightharpoonup}{x}}=\frac{1}{\rho}\ 
      \partial_{r} \frac{gg^{rs}}{\rho}\ \partial_{s} 
      \stackrel{\rightharpoonup}{x} = \displaystyle\sum_{j} 
      \{\{\stackrel{\rightharpoonup}{x}, x_{j}\}, x_{j}\}$ (defining
$\{f,g\} := \frac{1}{\rho}\  
      \epsilon^{rs} \partial_{r} f \partial_{s}g)$.
  \end{enumerate}
\end{enumerate}
Prove that a) and b), together with $G_{or}=0$, automatically imply 
the ``remaining'' equation $\triangle \zeta =0$! (So that $\zeta = 
x^0-x^3$, up to a constant, and provided $\epsilon^{rs} \partial_{r} 
\dot{\stackrel{\rightharpoonup}{x}} \partial_{s} 
\stackrel{\rightharpoonup}{x} \equiv 0$, can simply be thought of 
as being determined, via $G_{or}=0$, in terms of $x^1$ and $x^2$, 
satisfying b).) \bigskip

{\bf Excercise 2)} Consider now the case $d=2$ ($4-$dimensional 
Minkowski-space) \\
a) Show that the equations of motion 1b can be written as

\begin{eqnarray*}
\stackrel {\cdot}{x}_{1}&=&p_{1}, \hspace{4ex} \stackrel {\cdot}{p}_{1} =
\{\{x_{1},x_{2}\}x_{2}\} \\
\stackrel{\cdot}{x} _{2}&=&p_{2}, \hspace{4ex} \stackrel {\cdot}{p}_{2} =
\{\{x_{2},x_{1}\},x_{1}\}, where
\end{eqnarray*}

\[
\{f_{\prime}g\}:=\frac{1}{\rho}\in^{rs}\partial_{r}f\partial_{s}g
\]

b) Changing independent variables from $t(=\varphi^{0}), \varphi^{1},
\varphi^{2}$ to $x^{0}:=t, x^{1}, x^{2}$ show that $\{f,x_{r}\}=
-J\in_{rs} 
\frac {\partial \hat{f}}{\partial x^{s}}$, where $J(x)$ and $\hat{f}(x)$
are 
$\{x_{1},x_{2}\}$ resp. $f(\varphi)$ when expressed in the new
variables. 
Verify that the constraint $\in^{rs}\partial_{r}\stackrel{\stackrel
{\cdot}{\rightharpoonup}}{x} \partial_{s} \stackrel {\rightharpoonup}{x}$
therefore becomes the condition $\partial_{x_{1}} p_{2} - 
\partial_{x_{2}}p_{1}=0$, which can easily be solved: $\stackrel
{\rightharpoonup}{p} 
= \stackrel{\rightharpoonup}{\bigtriangledown} p(x)$

c) Calculate $\frac {\partial}{\partial \varphi^{0}} \{x_{1},x_{2}\}$ in 
2 different ways (via $\{p_{1},x_{2}\}+\{x_{1},p_{2}\}$, as well as
using 
$\partial_{\varphi_{0}} = \partial_{x_{0}} + \stackrel
{\rightharpoonup}{p}\cdot 
\stackrel {\rightharpoonup}{\bigtriangledown}$) to obtain the continuity 
equation
\[
\stackrel{\cdot}{q} + \stackrel {\rightharpoonup}{\bigtriangledown}
(q \stackrel {\rightharpoonup}{\bigtriangledown} p)=0
\] 
for the `velocity-potential' $p$ and the `density' $q:=\frac{1}{J}$

d) Convert the equations involving $\stackrel {\cdot}{p}_{r}$ (cp. $2a$) 
into `Euler's equation',
\[
\stackrel{\cdot}{p}=\frac {1}{2}(\frac{1}{q^{2}}-(\stackrel 
{\rightharpoonup}
{\bigtriangledown} p)^{2}+c
\]
for an irrotational, inviscid, isentropic gas whose presure is 
$\frac{-1}{\text{density}}$.

e) Choosing the constant $c$ to be zero, verify that the equations
obtained 
(cp. $2c,d$) are the ones that follow from $H=\frac{1}{2}\int d^{2}x
(q(\stackrel {\rightharpoonup}{\bigtriangledown}p)^{2}+\frac{1}{q})$
when 
$q$ and $p$ are treated as (hence ARE) canonically conjugate. \\

\pagebreak

\begin{center}
{\bf II}
\end{center}

Consider the theory of a time-dependent $M-$dimensional extended object 
$\Sigma_{M}(t)$ in $D=d+2$ dimensional Minkowski-space, defined by
requiring 
the Volume-functional 

\begin{eqnarray}
S[x^{\mu}]: &=& -\int d\varphi ^{0} d^{M} \varphi \sqrt{G} = -Vol 
(\mathcal{M})
          \label {c1} \\
          G &=& (-)^{M} \det 
          (\underbrace {\frac{\partial x ^{\mu}}{\partial \varphi 
^{\alpha}}
           \frac{\partial x ^{\nu}}{\partial \varphi ^{\beta}} \eta 
_{\mu \nu}}_{G _{\alpha \beta}})
          _{\alpha,\beta=0,1,\ldots ,M} 
          \nonumber
\end{eqnarray}

\noindent to be stationary under small variations of the embedding 
$-$functions
$x^\mu (\varphi ^{0}, \varphi ^{1},\ldots ,\varphi^{M}), \mu = 0, 1, 
\dots , d+1$
(describing the $M+1$ dimensional world volume $\mathcal{M}$).
(\ref{c1}) is invariant under diffemorphisms 
$\varphi^{\alpha} \rightarrow 
\tilde{\varphi}^{\alpha}(\varphi^{\beta})$, 
and inhomogeneous Lorentztransformations

\[
x^{\mu} \rightarrow \tilde{x}^{\mu} = \Lambda^{\mu}  _{\nu} x^{\nu} +
d^{\mu},\hspace{2ex} \Lambda ^{\mu}_{\mu^{\prime}} 
\Lambda^{\nu}_{\nu^{\prime}}
\eta_{\mu \nu} = \eta_{\mu^{\prime} \nu^{\prime}} 
=\mbox{diag} (1,-1,\ldots ,-1).
\]

Choosing
\begin{equation}
\varphi^{0}=\frac {x^{0}+x^{d+1}} {2} =:\tau
\label{c2}
\end{equation}

\noindent while denoting $x^{0}-x^{d+1}$ by $\zeta$,
differentiation with respect to $\tau$ by $\cdot$,
$(x^{1},\ldots,x^{d})$ by 
$\stackrel {\rightharpoonup}{x}$, and $G_{0r}=\partial_{r}\zeta - 
\stackrel{\stackrel{\cdot}{\rightharpoonup}}{x}
\partial_{r} \stackrel {\rightharpoonup}{x}$ by $u_r$,
the metric (induced from $\mathbb{R}^{1,d+1})$ on $\mathcal{M}$ reads 

\begin{equation}
(G_{\alpha \beta}) =
\left( \begin{array} {cc}
2\dot\zeta-\stackrel{\stackrel{\cdot}{\rightharpoonup}^2}{x} & u_{1}\ldots 
u_{M}
 \\
{u_{1}} & \\
\cdot & \\
\cdot & -g_{rs} \\
\cdot & \\
u_{M} & 
\end{array} 
\right)
\label{c3}       
\end{equation}

\[
g_{rs}=\partial_{r}\stackrel {\rightharpoonup}{x} \cdot \partial_{s} 
\stackrel {\rightharpoonup}{x}, 
\]

hence

\begin{equation}
G=g
(2 \stackrel {\cdot}{\zeta}-\stackrel{\stackrel{\cdot}
{\rightharpoonup}}{x}^{2}+u_{r} u_{s} g^{rs})
=:g \cdot \Gamma 
\label{c4}
\end{equation}   

with 
\[
g=\det(g_{rs})_{r,s=1\ldots M},
g^{rs} g_{s u}= \delta^{r}_{u}.
\]

So 
\begin{eqnarray}
S=S[\stackrel {\rightharpoonup}{x}, \zeta] &=& \int d\tau d^{M}\varphi 
\mathcal{L}
\nonumber \\
\mathcal{L}&=&-\sqrt{g}\sqrt{\Gamma}
\label{c5}
\end{eqnarray}

\noindent can be thought of as the action, resp. Lagrange density, of an
ordinary field
theory involving the $d+1$ fields $x^{1},\ldots,x^{d}, \zeta$.
Defining canonical momenta,

\begin{eqnarray}
\stackrel {\rightharpoonup}{p}: &=& \frac {\delta \mathcal{L}} 
{\delta \stackrel{\stackrel{\cdot}{\rightharpoonup}}{x}} = \sqrt 
{g/\Gamma}
(\stackrel{\stackrel{\cdot}{\rightharpoonup}}{x}+\partial_{r} 
\stackrel {\rightharpoonup}{x} u_{s} g^{rs})
\nonumber\\
\pi:&=& \frac {\delta \mathcal{L}} {\delta \stackrel {\cdot}{\zeta}} 
= - \sqrt{g/\Gamma}
\label{c6}
\end{eqnarray}

\noindent the Hamiltonian density reads

\begin{eqnarray}
\mathcal{H}: &=& \Pi \stackrel {\cdot}{\zeta} + \stackrel
{\rightharpoonup}{p} \cdot 
\stackrel{\stackrel{\cdot}{\rightharpoonup}}{x} + \sqrt{g} \sqrt{\Gamma}
\nonumber \\ 
&=& \sqrt{g/\Gamma} \{ \Gamma + 
\stackrel{\stackrel{\cdot}{\rightharpoonup}}
{x}^{2} 
+ \stackrel{\stackrel{\cdot}{\rightharpoonup}}{x} \partial_{r}
\stackrel{\rightharpoonup}{x} u_{s}g^{rs} - \stackrel {\cdot}{\zeta} \}
\label{c7} \\
&=& \sqrt {g/\Gamma} \{\stackrel {\cdot}{\zeta} + \partial_{r}\zeta
u_{s} g^{rs} \}.
\nonumber
\end{eqnarray}

Instead of now trying to eliminate the velocities in favour of the
momenta, 
simply note that (\ref{c7}) is the same as
 
\begin{equation}
\mathcal{H} =\frac {\stackrel {\rightharpoonup}{p}^{2}+g}{-2 \pi}
\label{c8}
\end{equation}

\[
(=1/2 \sqrt {g/\Gamma} \{ (\stackrel 
{\stackrel{\cdot}{\rightharpoonup}}{x}
+ \partial_{r} \stackrel {\rightharpoonup}{x} u_{s} g^{rs})^{2} + \Gamma\}
=\ldots=(\ref{c7})).
\]

As $\mathcal{H}$ does not contain $\zeta$ (!), the classical equation of 
motion
$\stackrel {\cdot}{\pi}=-\frac{\delta \int \mathcal{H}}{\delta\zeta} =
0$ imply that $\pi$ 
is a constant (density),

\begin{equation}
\pi= -\eta \rho(\varphi), \hspace{5ex} \int d^{M}\!\varphi \rho(\varphi)
= 1,
\label{c9}
\end{equation}

\noindent so that the Hamiltonian

\begin{equation}
H_{(-)} = \int \mathcal{H} = \frac {1}{2 \eta} \int_{\sum^{(M)}} 
d^{M}\!\varphi 
\frac {\stackrel {\rightharpoonup}{p}^{2}+g} {\rho(\varphi)} 
\label{c10}
\end{equation}

\noindent becomes polynomial in the dynamical fields $(\stackrel
{\rightharpoonup} {x}$ 
and $\stackrel {\rightharpoonup} {p})$- which is the crucial difference
to all 
the Hamiltonian formulations mentioned in section {\bf I} (they all 
contain square-roots). 
But what has happened to $\zeta$?! (After all, as it is part of the 
description of the manifold $\mathcal{M}$, it can't just disappear
altogether).
The point is that actually the transition to (\ref{c10}) is not
quite as straightforward as pretended, due to the presence of the
constraints \footnote{just insert (\ref{c6}), to verify (\ref{c11})} 
(reflecting the remaining time-dependent spatial reparametrisation 
invariance after the partial gauge-fixing (\ref{c2}))

\begin{eqnarray}
\Phi_{r}:=\pi \partial_{r}\zeta+ \stackrel{\rightharpoonup}{p} \partial 
_{r}
\stackrel{\rightharpoonup}{x} \equiv 0
\label{c11} \\
r=1,\ldots,M \nonumber ;
\end{eqnarray}

\noindent (so strictly speaking (\ref{c10}) is only defined up to terms
$\int {d^{M}\!\varphi} u^{r} \Phi_{r}$; putting them to zero, as done in
(\ref{c10}), amounts to the further gange choice $G_{0r}=u_{r}=0$
(cp. the excercise at the end of the section I) - which explains the 
further reduction of the invariance group to volume-preserving 
Diffeomorphisms of $\sum_{M}$ in (\ref{c10}). 
On the other hand, (\ref{c11}) precisely resolves the puzzle of 
the missing $\zeta$ as, given (\ref{c9}), it allows to determine $\zeta$
(up to a constant which is canonically conjugate to $\eta$) in terms 
of the dynamical fields $\stackrel {\rightharpoonup}{x}$, as long as the 
integrability
conditions

\begin{equation}
\Phi_{rs}:=\partial_{r} (\stackrel {\rightharpoonup}{p}/_\rho) \partial
_{s}\stackrel {\rightharpoonup}{x} - 
\partial_{s}(\stackrel {\rightharpoonup}{p}/_\rho) \partial_{r}
\stackrel {\rightharpoonup}{x}   \equiv 0 
\label{c12} 
\end{equation}

\noindent hold (-which are consistent with (\ref{c10}), as the
$\Phi_{rs}$ are the 
generators of the above mentioned symmetry group $SDiff \sum _{M}$ of 
(\ref{c10})).
Finally noting that $H_{(-)}$ is generating translations in $\tau$ (just 
as
$\stackrel {\rightharpoonup}{P}: = \int {\stackrel
{\rightharpoonup}{p}}$ 
generates translations in 
$\stackrel {\rightharpoonup}{x}$, and $P_{+}:=-\int \pi d^{M} {\varphi}
= 
\eta$ in $x^{0}-x^{3}$), 
one sees that the relativistically invariant 

\[
\mathbb{M}^{2}:=P^{\mu} P_{\mu} \equiv 2P_{+}P_{-} -\stackrel 
{\rightharpoonup}{P}^{2}
\]

\noindent takes the simple form

\begin{equation}
\mathbb{M}^{2}=\int \frac {d^{M} \varphi}{\rho(\varphi)} 
(\stackrel {\rightharpoonup}{p}^{2}+g)
-\stackrel {\rightharpoonup}{P}^{2}
\label{c13}
\end{equation}

\noindent which may be rewritten in the following ways: \\
first of all,
\begin{equation}
g=\sum_{i_{1}<i_{2}<\ldots<i_{M}} 
\{x_{{i}_{1}},x_{{i}_{2}},\ldots,x_{{i}_{M}}\}^{2}\ ,
\label{c14}
\end{equation}

\noindent where the multilinear bracket is defined, for any set of $M$
differentiable 
functions on $\sum_{M}$, as

\begin{equation}
\{x_{{i}_{1}},\ldots,x_{{i}_{M}}\}:=\frac{1}{\rho}\in^{r_{1}\ldots r_{M}}
\frac{\partial x_{i_1}}{\partial \varphi ^{r_1}} \ldots
\frac{\partial x_{i_M}}{\partial \varphi ^{r_M}}\ .
\label{c15}
\end{equation}

Secondly, expanding the fields $x_{i}(t,
\varphi^{1},\ldots,\varphi^{M})$ and
their conjugate momenta $p_{i}=\rho p_{i\alpha}Y_{\alpha}$ in terms of 
basis-
functions $Y_{\alpha}, \alpha=0,1,\ldots,$ on $\sum_{M}, \int
{d^{M}\varphi} 
\rho Y_{\alpha} Y_{\beta} = \delta_{\alpha \beta}$, (\ref{c13}) becomes 

\begin{equation}
\mathbb{M}^{2}=p_{i\alpha} p_{i\alpha}+\frac {1}{M!} 
g_{\alpha \alpha_1 \ldots \alpha_M}
g_{\alpha\beta_1\ldots\beta_M}
x_{i_{1}\alpha_1\ldots}
x_{i_{M} \alpha_M}
x_{i_1\beta_1\ldots}
x_{i_M \beta_M}\ ,
\label{c16}
\end{equation} 

\noindent where the `structure-constants'

\begin{equation}
g_{\alpha \alpha_{1}\ldots\alpha_{M}}:=\int d^{M}\!\varphi \rho Y_{\alpha}
\{Y_{\alpha_1},\ldots,Y_{\alpha_{M}}\}
\label{c17}
\end{equation}

\noindent encode all the information about the fact that one is dealing
with a 
certain $M-$dimensional object (a sphere $S^{M}$ or a torus $T^{M}$, 
or $\dots$); the sum over $\alpha, \beta$ is defined to run only from $1$
to $\infty$ (rather than $0$ to $\infty$; due to the 
$\stackrel {\rightharpoonup}{P}^{2}$
subtraction, and $\{.,\ldots ,.\}$ containing only derivatives, 
all zeromodes drop out). The potential
$\mathbb{M}^{2}-p_{i\alpha}p_{i\alpha}$ is a 
homogeneous polynomial of degree $2M$, with the notable property that in 
each term, each coordinate appears at most quadratically. While for $M>2$
the question of how to quantize (\ref{c16}) is mostly open (see [23]),
let us
come back to the original case of interest, to membranes; we then have 
(dropping the factor $\frac{1}{2\eta}$ in (\ref{c10}), resp. not making
any 
notational distinction between $H_{(-)}$ and $\mathbb{M}^{2}$,

\begin{eqnarray}
H &=& p_{i \alpha}p_{i \alpha} + \frac{1}{2} g_{\alpha \beta \gamma} 
g_{\alpha, \beta^{\prime} \gamma^{\prime}} 
x_{i\beta} x_{j\gamma} x_{i\beta^{\prime}} x_{j\gamma^{\prime}}
\label{c18} \\
&=& \int _{\sum_{2}} d^{2}\varphi \rho
((\frac {\stackrel {\rightharpoonup}{p}} {\rho})^{2} + 
\sum_{i<j}\{x_{i},x_{j}\}^{2}) -\stackrel {\rightharpoonup}{P}^{2}
\nonumber
\end{eqnarray}

\noindent with

\begin{equation}
g_{\alpha \beta \gamma}:= 
\int _{\sum_{2}} d^{2}\varphi Y_{\alpha} 
(\frac {\partial Y_{\beta}} {\partial \varphi^{1}} 
\frac {\partial Y_{\gamma}} {\partial \varphi^{2}} - 
\frac {\partial Y_{\gamma}} {\partial \varphi^{1}}
\frac {\partial Y_{\beta}} {\partial \varphi^{2}})
\label{c19}
\end{equation}

\noindent being structure constants (in the basis corresponding to the 
$Y_{\alpha}$)
of the infinite-dimensional Lie algebra $s diff \sum_{(2)}$ of 
divergence-free vectorfields (resp. functions, modulo constants) on the 
space $\sum_{2}$ (parametrizing the surface).
(\ref{c18}) has to be supplemented by the constraints 
$(\alpha=1,2,\ldots)$

\begin{equation}
g_{\alpha \beta \gamma} x_{j_{\beta}} p_{j_{\gamma}} = 0\ .
\label{c20}
\end{equation}

$SO(d+1,1)$ - invariance of this (classical) theory 
was proven by Goldstone [24].

Formally, one could now try to define a `quantum theory of relativistic 
surfaces' by putting `hats' on all $x$'s and $p$'s, and demanding 
canonical
commutation relations, 
\[
[\hat{x}_{j_{\alpha}}, \hat{p}_{k_{\beta}}]= i\not{h} \delta_{jk}\delta
_{\alpha \beta}\ ;
\]

\noindent representing $\hat{p}_{k\beta}$ by $-i\hbar\ \frac
{\partial}{\partial 
x_{k_{\beta}}}$, 
one would thus arive at an (a priori ill-defined) infinite-dimensional 
Schr\"{o}dinger operator
\[
-\Delta_{\infty} + V_{\infty}\ .
\]
Fortunately, there exists a (symmetry maximally preserving) regulation 
procedure \footnote{originally discovered in the case of a 2-sphere [1], 
10
years later realized for $T^{2}$ [25], conjectured to hold for higher
genus 
surfaces [26], and finally proven, for general K\"{a}hler manifolds,
in [27].} 
amounting to the following

{\bf Theorem:}

For each $\sum_{2}$ there exists a basis $ \{ 
Y_{\alpha}\}^{\infty}_{\alpha=1}$ of $sdiff \sum_{2}$, and a basis
$\{T_a^{(N)}\}^{N^{2}-1}_{a=1}$ of
$su(N)$ such that

\begin{equation}
\lim_{N\rightarrow \infty} Tr
(-\underbrace {[T_{a}^{(N)},T_{b}^{(N)}] T_{c}^{(N)}}_{=f_{abc}^{(N)}}) =
g_{abc} \hspace{1ex}
\forall_{a,b,c}\ . \label{c21}
\end{equation}

One could therefore consider the class of $SU(N)$ invariant matrix
models 
$(X_{i}:=x_{ia}T_{a}^{(N)})$ 

\begin{equation}
H_{N}:= \sum ^{d}_{i=1} \sum ^{N^{2}-1}_{a=1}p_{ia} p_{ia}+ \frac{1}{2}\
f^{(N)}_{abc}
f^{(N)}_{ab'c'}
x_{ib}x_{jc}x_{ib^{\prime}}x_{jc^{\prime}}
\label{c22}
\end{equation}

\noindent as approximating (\ref{c18}), with 

\begin{equation}
f_{abc}^{(N)} x_{jb} p_{jc}=0
\label{c23}
\end{equation} 

\noindent replacing (\ref{c20}). These finite-dimensional models can be 
quantized without 
any problem, thus arriving at ([1])

\begin{equation}
\hat{H}_{N}=-\Delta_{(N)} - Tr \sum_{i<j}[X_{i},X_{j}]^{2}\ ,
\label{c24}
\end{equation} 

\noindent $N\rightarrow\infty$, as a quantized discrete analogue of a
relativistic 
membrane theory in $\mathbb{R}^{1,d+1}$. 

$\hat{H}_N$ commutes with the operators

\begin{eqnarray}
\hat{K}_{a}^{(N)}:=-f_{abc}^{(N)} x_{jb} \frac{\partial}{\partial x_{jc}}
\label{c25} \\
a=1\ldots N^{2}-1 \nonumber
\end{eqnarray} 

\noindent while $[\hat{K}_{a}^{(N)},\hat{K}_{b}^{(N)}]=i f_{abc}^{(N)} 
\hat{K}_{c}^{(N)}$; 
it is therefore consistent to restict to square-integrable warefunctions 
$\psi(x_{ja})$ that are annihilated by (\ref{c25})$^{N^{2}-1}_{a=1}$,
and because of (\ref{c23}), this {\em is} the physical Hilbert-space to
be 
considered. \\
\hspace*{2ex} Let me now prove (\ref{c21}) for $S^{2}$ and $T^{2}$: In
the case of the 
$2$-sphere the basis of $su(N)$ is obtained by considering the usual
spherical 
harmonics $\{Y_{lm}(\theta,\varphi)_{lm}\} _{\stackrel {l=1 \ldots 
\infty}{m=-l\ldots +l}}$ 
writing them as harmonic homogeneous polynominals in 
$x_{1}=r\sin\theta\cos\varphi, x_{2}=r\sin\theta\sin\varphi, 
x_{3}=r\cos\varphi$
(restricted to $r^{2}=\stackrel {\rightharpoonup}{x}^{2}=1$):

\begin{equation}
Y_{lm} {(\theta,\varphi)}=\sum c^{(m)}_{a_{1}\ldots a_{l}} 
x_{a_{1} \cdot \ldots \cdot} x_{a_{l}}|_{\stackrel 
{\rightharpoonup}{x}^{2}=1}
\label{c26} 
\end{equation} 

\noindent (where the tensor $c\ldots$ is by definition traceless and 
totally symmetric)
and then replacing the commuting varibles $x_a$ by generators $X_a$ of 
the $N-$dimensional irreducible (spin $s=\frac{N-1}{2})$ representation
of 
$su(2)$, to obtain $N^{2}\!-\!1$ $N\times N$ matrices defined by [1]

\begin{eqnarray}
T_{lm}^{(N)}:&=&\gamma_{Nl} \sum c^{(m)}_{a_{1}\ldots a_{l}} 
X_{a_{1}} \cdot \ldots \cdot X_{a_{l}}; 
\label{c27} \\
l &=& 1\ldots N^{2}-1 \nonumber \\
m &=& -l,\ldots,+l \nonumber
\end{eqnarray} 

\noindent instead of having $X^{2}_{1}+ X^{2}_{2}+ X^{2}_{3}
=\frac{N^{2}-1}{4}\ \unity$, which makes the explicit proof of
(\ref{c21}) 
quite involved, it is easier to choose $\stackrel {\rightharpoonup}{X}
^{2}= \unity$, i.e.

\begin{equation}
[X_{a},X_{b}]= \frac{2i}{\sqrt{N^2-1}}\ \epsilon_{abc} X_{c}
\label{c28}
\end{equation}

\noindent and
\[
\gamma_{Nl}= -i\sqrt{\frac{N^{2}-1}{4}}.
\]

As the Poisson-bracket $\{,\}$ on $S^{2}$ can be thought of as coming
from 

\begin{equation}
\{f (\stackrel {\rightharpoonup}{x}), g {(\stackrel 
{\rightharpoonup}{x})}\}_{\mathbb{R}^{3}}:=
\in_{abc} \partial_{a}f \partial_{b}g\ x_{c}\ ,
\label{c29}
\end{equation}

\noindent $\{Y_{lm},Y_{l^{\prime}m^{\prime}}\}$ can be computed from 
$\{r^{l}Y_{lm},r^{l\prime}Y_{l\prime m\prime}\}_{\mathbb{R}^{3}}$

\begin{eqnarray}
=\sum c_{a_{1} \ldots a_{l}}^{(m)}\ c_{b_{1} \ldots b_{l'}}^{(m')}
\{x_{a_{1}}\dots x_{a_{l}},x_{b_{1}}\ldots x_{b_{l'}}\}_{{\mathbb{R}}^3}
\nonumber 
\end{eqnarray}
by using the derivation proporty of $\{\cdot , \cdot \}$, and

\begin{equation} 
\{x_{a},x_{b}\}= \in_{abc} x_{c}
\label{c31},
\end{equation}

\noindent as well as then decomposing the resulting polynomial of degree
$l+l^{\prime}-1$ 
into harmonic homogeneous ones.
Calculating

\begin{eqnarray}
[T^{(N)}_{lm}, T^{(N)}_{l'm'}]
=-\frac {\stackrel {2}{N-1}}{4} 
\sum c^{(m)}_{a_{1}\dots a_{l}}\ c^{(m')}_{b_{1}\dots b_{l'}}
[X_{a_{1}}\dots X_{a_{l}}\ X_{b_{1}} \dots X_{b_{l'}}]\ , \label{c32}
\end{eqnarray}

\noindent the first step is identical to the above, while any further
use of the 
commutation-relations (\ref{c28}), - necessary to obtain the desired
traceless 
totally symmetric tensors -, introduces factors of 
$\frac{1}{\sqrt{N^{2}-1}}$; hence the agreement of $f^{(N)}_{abc}$
and $g_{abc}$ to leading order in N. Concrete expressions for the matrix
elements 
of the $T^{(N)}_{lm}$ in terms of hypergeometric functions, and related
discrete 
orthogonal polynomials, are given in [28]. 

For the $2$--torus, things are even easier. Taking 
  $Y_{\stackrel{\rightharpoonup}{m}} := - 
  e^{i(m_{1}\varphi_{1}+m_{2}\varphi_{2})},$ \linebreak $(m_{1}, m_{2})
= 
  \stackrel{\rightharpoonup}{m} \in {\mathbb{Z}}^2$, the relevant basis
of 
  $su(N), N$ odd, is given by (cp. [25])
  {\def\theequation{54}
   \begin{equation}
       T_{\stackrel{\rightharpoonup}{m}}^{(N)} := \frac{iN}{4\pi M}\ 
       \omega^{\frac{1}{2}\,m_{1}m_{2}} g^{m_{1}} h^{m_{2}}
    \end{equation}}
  where
  {\def\theequation{55}
   \begin{eqnarray}
  \omega & = &e^{\frac{4\pi iM}{N}}\ ,  \\ g & = &\left( 
 \displaystyle \begin{array}{ccccc}
      1 & & & & \\
      & \omega & & & 0 \\
      & & \omega^2 & & \\
      & & & \ddots & \\
      0 & & & & \omega^{N-1}
      \end{array} \right)\ , \hspace{2ex} h \hspace{1ex} = 
      \hspace{1ex} \left( 
  \begin{array}{cccc}
      0 & 1 & &  \\
      & \ddots & \ddots \\
      & & \ddots & 1 \\
      1 & & & 0
      \end{array} \right) , \nonumber
  \end{eqnarray}}
  and $M \in {\mathbb{N}}$ having no common divisor with $N$. Using the
basic 
  relation $h \cdot g = \omega g \cdot h$ it is very easy to verify 
  that
  {\def\theequation{56}
  \begin{equation}
     [T_{\stackrel{\rightharpoonup}{m}}^{(N)}, 
     T_{\stackrel{\rightharpoonup}{n}}^{(N)}] = \frac{N}{2\pi M}\ 
     \sin\ (\frac{2\pi M}{N}\ (\stackrel{\rightharpoonup}{m} \times 
     \stackrel{\rightharpoonup}{n}))\ 
    T_{\stackrel{\rightharpoonup}{m}+\stackrel{\rightharpoonup}{n}}^{(N)}\ 
.
     \end{equation}}
  The trigonometric structure constants (of $g\ell (n, {\mathbb{C}})$ in
the 
  basis \linebreak $\{T_{\stackrel{\rightharpoonup}{m}}^{(N)}\}_{m_{1},
m_{2}= - 
  \frac{(N-1)}{2}, \dots , + \frac{(N-1)}{2}})$ indeed converge (for 
  $N \to \infty,\ M$ fixed) to $\stackrel{\rightharpoonup}{m}\times 
  \stackrel{\rightharpoonup}{n} = m_{1} n_{2} - m_{2} n_{1}$, i.e. 
  those of $s$\,diff\,$T^2$ in the basis 
  $\{-e^{i \stackrel{\rightharpoonup}{m} 
  \stackrel{\rightharpoonup}{\varphi}}\}_{\stackrel{\rightharpoonup}{m}
\in {\mathbb{Z}}^2}$. 
  To demonstrate the 
  subtleness of this (non--inductive) limit note that if $M = M(N)$ is 
  chosen such that $\displaystyle\lim_{N\to\infty} \frac{M(N)}{N} = 
  \Lambda \in [0, \frac{1}{4})$, one obtains infinite dimensional 
  Lie--Algebras $L_{\Lambda}$ with commutation--relations
  \[
  [T_{\stackrel{\rightharpoonup}{m}}, 
  T_{\stackrel{\rightharpoonup}{n}}] = \frac{1}{2\pi \Lambda}\ \sin\ (2 
  \pi \Lambda  ( \stackrel{\rightharpoonup}{m} \times 
  \stackrel{\rightharpoonup}{n}))\ T_{\stackrel{\rightharpoonup}{m}+
  \stackrel{\rightharpoonup}{n}}\ , 
  \]
  which for irrational $\Lambda$ are all non--isomorphic (see e.g. 
  [39]) and (up to $T_{\stackrel{\rightharpoonup}{0}}$) simple 
  (for rational $\Lambda,\ L_{\Lambda = 
  p/q}$ contains a large ideal $I$ of finite codimension, with 
  $L_{\Lambda}/_{I} \cong g\ell (N, {\mathbb{C}})$).
  
  \newpage

\subsection*{Excercise 3:} (cp. [22], [28])

\smallskip

\noindent Assuming that the $T_{\ell m}^{(N)}$, like the $Y_{\ell m}$, 
transform as spherical tensors, leading to orthogonality relations
$$
Tr \, T_{\ell m}^{\dagger} \, T_{\ell' m'} = \delta_{\ell \ell'} \, 
\delta_{mm'} 
\, 
r_N^2 (\ell) \, ,
$$
calculate $r_N (\ell)$ (from (49)/(50)) by using that
$$
r^{\ell} \, Y_{\ell\ell} = (-)^{\ell} \sqrt{\frac{2\ell + 1}{4\pi}} \, 
\sqrt{\frac{(2\ell)!}{2^{2\ell} (\ell!)^2}} \, (x_1 + ix_2)^{\ell} \, ,
$$
i.e.
\begin{eqnarray}
r_N^2 (\ell) &= &\frac{2\ell + 1}{4\pi} \, \frac{(2\ell)!}{2^{2\ell} 
(\ell!)^2} \, \frac{N^2 - 1}{4} \, Tr ((X_1 + i X_2)^{\ell} (X_1 - i 
X_2)^{\ell}) \nonumber \\
&= &\cdots = \frac{(N+\ell)!}{16 \pi (N-\ell - 1)! \, (N^2 - 1)^{\ell - 1}} 
\nonumber
\end{eqnarray}
(in quite a few ``fuzzy-sphere-articles'' this non-trivial $\ell$-dependence 
has been forgotten).

\bigskip

\subsection*{Excercise 4:} (cp. [43])

\smallskip

In terms of hermitian $N \times N$ matrices $X_j$, the classical equations of 
motion corresponding to (44)/(45) are
$$
\ddot X_i = - \sum_{j=1}^d \ [[X_i , X_j] , X_j]
$$
$$
\sum_{j=1}^d \ [X_j , \dot X_j] = 0 \, .
$$
Verify the following types of solutions:

\medskip

\noindent A)
\begin{eqnarray}
X_i (t) &= &x(t) \, r_{ij} (t) \, M_j \nonumber \\
&= &x(t) \, (\cos \varphi (t) \cdot \stackrel{\rightharpoonup}{M} , \, \sin 
\varphi (t) \cdot \stackrel{\rightharpoonup}{M} , 0 \ldots 0) \nonumber
\end{eqnarray}
with
$$
\frac{1}{2} \, \dot x^2 + \frac{\lambda}{4} \, x^4 + \frac{L^2}{2} \, 
\frac{1}{x^2} = {\rm const} \qquad\qquad\qquad x^2 \, \dot{\varphi} = L \, ,
$$
$$
\sum_{a=1}^{\widetilde d} \ [[M_a , M_b],M_b] = \lambda \, M_a \qquad (*)
$$
and

\smallskip

\noindent I) 
$$
[M_a , M_b] = f_{abc} \, M_c, \, f_{abc} \, f_{cba'} = - \lambda \, f_{aa'}
$$

\noindent II) 
$$
\stackrel{\rightharpoonup}{M} = \frac{1}{\sqrt 2} \left( \frac{g+g^{-1}}{2} , 
\frac{g-g^{-1}}{2i} , \frac{h+h^{-1}}{2} , \frac{h-h^{-1}}{2i} , 0 \ldots 0 
\right)
$$
($g$ and $h$ as in (55)).

\smallskip

\noindent III)
\begin{eqnarray}
M_1 &= &\frac{\pi M}{iN} \, (T_{k\ell} + T_{-k-\ell} + T_{-k\ell} + 
T_{k-\ell}) \nonumber \\
M_2 &= &\frac{-\pi M}{N} \, (T_{k\ell} - T_{-k-\ell} - T_{-k\ell} + 
T_{k-\ell}) \nonumber \\
M_3 &= &\frac{-\pi M}{N} \, (T_{k\ell} - T_{-k-\ell} + T_{-k\ell} - 
T_{k-\ell}) \nonumber \\
M_4 &= &\frac{-\pi M}{iN} \, (T_{k\ell} + T_{-k-\ell} - T_{-k\ell} - 
T_{k-\ell}) \nonumber
\end{eqnarray}
($T_{k\ell}$ as in (54)).

\medskip

\noindent B)
$$
X_j (t) = \sum_{\alpha} r_{\alpha} (t) ( 
\stackrel{\rightharpoonup}{E}_{\alpha})_j
$$
with
$$
\stackrel{\rightharpoonup}{E}_1 = \frac{2\pi M}{iN} 
(T_{\stackrel{\rightharpoonup}{m}_1} + T_{-\stackrel{\rightharpoonup}{m}_1} , 
-i (T_{\stackrel{\rightharpoonup}{m}_1} - 
T_{-\stackrel{\rightharpoonup}{m}_1}) , 0 \ldots 0)
$$
\begin{eqnarray}
\stackrel{\rightharpoonup}{E}_2 &= &\frac{2\pi M}{iN} (0,0, 
T_{\stackrel{\rightharpoonup}{m}_2} + T_{-\stackrel{\rightharpoonup}{m}_2} , 
-i (T_{\stackrel{\rightharpoonup}{m}_2} - 
T_{-\stackrel{\rightharpoonup}{m}_2}) , 0 \ldots 0) \nonumber \\
&&\vdots \nonumber \\
&&\ddot r_{\alpha} (t) = - 4 \, r_{\alpha} \sum_{\beta} \sin^2 \left( 
\frac{2\pi M}{N} \, (\stackrel{\rightharpoonup}{m}_{\alpha} \times 
\stackrel{\rightharpoonup}{m}_{\beta}) \right) r_{\beta}^2 \, .\nonumber
\end{eqnarray}
  
  \newpage
  
  \begin{center}
   {\bf  III}
\end{center}

Let me now turn to the supersymmetric extension [4, 5] of the bosonic 
matrix model (44) which, 
following [6], has been intensively studied over the past few years, 
as a candidate for 
`$M$--theory' 
{\renewcommand{\thefootnote}{\fnsymbol{footnote}}\footnote[1]{Sorry 
for the change of notation; $su(N)$ indices are now denoted by: 
$A, B, C = 1, \dots , N^2-1$, transverse space--time indices by 
$s, t = 1, \dots , d (= 2, 3, 
5$ or $9$).}} {\def\theequation{1}
\begin{equation}
H_{Susy} = p_{tA} p_{tA} + \frac{1}{2}\ f_{ABC} f_{AB'C'} q_{sB} q_{tC} 
q_{sB'} q_{tC'} + iq_{tC} f_{ABC} \gamma_{\alpha\beta}^t 
\Theta_{\alpha A} \Theta_{\beta B}\ ,
\end{equation}}
$\alpha, \beta = 1, \dots , s_{d} (= 2, 4, 8$ or 
$16);$ $f_{ABC} =$ structure constants of $SU(N)$ (real, 
antisymmetric), \\
$\gamma^s \gamma^t + \gamma^t 
\gamma^s = 2 \delta^{st} \unity_{s_{d}\times s_{d}},\ \gamma^s$ real, 
symmetric;

the dynamical degrees of freedom satisfy canonical 
(anti)commutation relations,
{\def\theequation{2}
\begin{equation}
   [q_{tA}, p_{sB}] = i\delta_{AB} \delta_{ts} \hspace{5ex} 
   \{\Theta_{\alpha A}, \Theta_{\beta B}\} = \delta_{\alpha\beta} 
   \delta_{AB}\ .
\end{equation}}
$H_{Susy}$ commutes with the generators of $SU(N), Spin(d)$ and 
supersymmetry,
{\def\theequation{\arabic{equation}}
\setcounter{equation}{2}
\begin{eqnarray}
    J_{A} & = & f_{ABC} (q_{sB} p_{sC} - \frac{1}{2}\ i \Theta_{\alpha 
    B} \Theta_{\alpha C}) \nonumber \\
    J_{st} & = & q_{sA} p_{tA}  - q_{tA} p_{sA} - \frac{1}{4}\ i 
    \Theta_{\alpha A} \gamma_{\alpha\beta}^{st} \Theta_{\beta A} 
    \\
    Q_{\beta} & = & (p_{tA} \gamma_{\beta\alpha}^t + \frac{1}{2}\ 
    f_{ABC} q_{sB} q_{tC} \gamma_{\beta\alpha}^{st}) \Theta_{\alpha 
    A} \nonumber
\end{eqnarray}}
while
{\def\theequation{4}
\begin{equation}
\{Q_{\beta} , Q_{\beta'}\} = \delta_{\beta\beta'} H + 
2\gamma_{\beta\beta'}^t q_{tA} J_{A}\ ,
\end{equation}}
and all states $\Psi$ are required to be $SU(N)$--invariant, 
$J_{A}\Psi = 0$ (so that, on the physical Hilbert space ${\mathcal{H}}, 
H_{Susy}$ is twice the square of each $Q_{\beta}$).

In verifying (4), $d=2, 3, 5$ and $9$ are singled out when 
calculating the mixed $(pqq)$ terms, which yield (only) the 
fermionic (last term) part of (1) and the `weakly zero'--part 
(proportional to the $J_{A}$) in (4) {\em provided}
{\def\theequation{5}
\begin{equation}
 \gamma_{\alpha\beta}^s \gamma_{\alpha'\beta'}^{st}  +
\gamma_{\alpha'\beta}^s 
 \gamma_{\alpha\beta'}^{st} + \gamma_{\alpha\beta'}^s 
 \gamma_{\alpha'\beta}^{st} + \gamma_{\alpha'\beta'}^s 
 \gamma_{\alpha\beta}^{st} = 2 (\delta_{\alpha\alpha'} 
 \gamma_{\beta\beta'}^t - \delta_{\beta\beta'} \gamma_{\alpha\alpha'}^t)
\end{equation}}
-- which (setting e.g. $\beta = \beta'$ and summing) implies
{\def\theequation{6}
\begin{equation}
    s_{d} = 2 (d-1)\ ;
\end{equation}}
together with the reality condition on the ${\gamma'}{s}$ (to have
$H_{Susy}$ 
and $Q_{\beta}$ hermitean) these hold only when $d = 2, 3, 5$ or $9$.

The spectrum of $H_{Susy} \ge 0$ is known to cover the whole positive 
axis ${\R}^+$ [8,29] and it is conjectured that for $d=9$ there exists 
a unique normalizable state for each $N$, while {\em not} for all 
other cases $(d=2, 3, 5)$. For $N=2$, the latter was proven in [30] 
and for $d=9$ the precise form of a (unique) asymptotic solution was 
derived in [31,32]. Writing [32]
{\def\theequation{7}
\begin{equation}
    q_{tA} = re_{A} E_{t} + y_{tA}
\end{equation}}
for configurations with almost vanishing potential energy $(V = 0$ 
for $y_{tA}=0; E_{t}E_{t}=1=e_{A}e_{A}, r \to \infty)$ the asymptotic 
wavefunction takes the form
{\def\theequation{8}
\begin{equation}
 \Psi \sim r^{-\kappa} e^{-ry^2/2} | F^{\bot}\rangle | F^{||}\rangle
\end{equation}}
with $\kappa = 0, -1, 6$ for $d = 3, 5, 9$ (respectively), $y^2 = 
y_{tA} y_{tA}, | F^{||}\rangle$ only involving $\Theta_{\alpha A} 
e_{A}$ and $E_{s}$, and $|F^\bot\rangle$ involving all other 
fermionic degrees of freedom (and $E_{s}$).

For $d=9, r^4 e^{-ry^2/2} | F^\bot\rangle$ corresponds to the ground 
state of a system of $16$ supersymmetric harmonic oscillators, and
{\def\theequation{9}
\begin{equation}
   |F^{||}\rangle = (E_{s} E_{t} - \frac{1}{9}\ \delta_{st}) |44; 
   st\rangle\ ,
\end{equation}}
where $|44; st\rangle$ is the only $spin (9)$ representation in the 
$256$--dimensional fermionic Hilbert--space ${\mathcal{H}}_{||} (= 44 
\oplus 84 \oplus 128$, arising from the $16$--dimensional spinor  
$\Theta_{\alpha}^{||} := e_{A} \Theta_{\alpha A}$) that can be
contracted with 
a bosonic $spin (9)$ representation (made out of the $E's$) to form a 
$spin (9)$--singlet. 

For $N>2$, the asymptotic wavefunction, in accordance with (8), is 
speculated to factorize into the ground state of a system of 
supersymmetric harmonic oscillators, times a supersymmetric spin $9$ 
(and Weyl!--)\, invariant wavefunction, $\Psi_{c}$, involving only the 
Cartan--subalgebra degrees of freedom, $\Theta_{\alpha k}$ and $q_{sk} 
\hspace{1ex} (k = 1, 2, \dots , N-1)$, and annihilated by effective 
(free) supercharges
{\def\theequation{10}
\begin{equation}
 {\Q}_{\beta} := - i\ \frac{\partial}{\partial q_{uk}}\ 
 \gamma_{\beta\alpha}^{u} \Theta_{\alpha k}\ .  
\end{equation}}
Unlike the $N=2$ case, where $r^{-9} | F^{||}\rangle \sim 
\partial_{s}\partial_{t} (\frac{1}{r^7}) |44; st\rangle$ was the 
\underline{only} $spin (9)$ invariant wavefunction\footnote{in order to
explicitely 
check that it is supersymmetric one may use that 
$\Theta_{\alpha}^{||}|44;st\rangle=\gamma_{\alpha\beta}^s|t\beta\rangle
+ 
\gamma_{\alpha\beta}^t |s\beta\rangle$, which leads to a fermionic 
$[us]$ resp. $[ut]$ antisymmetry, giving indeed zero, when contracted 
with $\partial_{s}\partial_{t}\partial_{u}\ (\frac{1}{r^7})$\ [35].},
there 
are abundantly many harmonic wavefunctions,
{\def\theequation{11}
\begin{equation}
 \Psi_{c} = \sum_{\ell,S,R,m} r^{-2\ell-(N-1) d+2-S\times R} \Psi_{\ell 
 m} (\stackrel{\rightharpoonup}{q}_{1}, \dots , 
 \stackrel{\rightharpoonup}{q}_{N-1}) | S \times R; m\rangle
\end{equation}}
for $N \ge 3\ [36,37];\ S$ and $R$ label the irreducible 
representations of the permutation group $S_{N}$ ($=$ the Weyl group 
of $SU(N)$) and $spin (9)$, $\overline{\Psi}_{\ell m}^{S\times R}$ is a 
harmonic (homogenuous of degree $\ell$) polynomial in the $d(N-1)$ 
bosonic variables $q_{tk}, | S \times R; m\rangle, m = 1 \dots 
\dim\, (S \times R)$, is a corresponding $S \times R$ representation 
made out of the fermions, and $r^{-2\ell - (N-1) d+2}$ is the 
(unique) power making $\Psi_{c}$ a decaying harmonic function. 
The (non--trivial) question then is: which of the wavefunctions (11) 
is annihilated by (10)?

For $N=3$, the answer was guessed in [37] and proven in [38]:
{\def\theequation{12}
\begin{equation}
    \Psi_{c} = \left(\prod_{\beta=1}^{16} {\Q}_{\beta}\right) (r^{-16} |
1 \times 
    1\rangle )\ .
\end{equation}}
Assuming that also for higher odd $N$ the fermionic Hilbertspace will 
contain a unique Weyl-- and $spin (9)$ invariant state, one would guess 
that
{\def\theequation{13}
\begin{equation}
    \Psi_{c} = \left(\prod_{\beta=1}^{16} {\Q}_{\beta}\right) (r^{-9
(N-1)+2} | 1 \times 
    1\rangle_{N} )
\end{equation}}
for odd $N \ge 3$ (but note that even for $N=3$ the proof that (12) 
is non--vanishing does, a priori\footnote{the possibility that the 
uniqueness of a normalizable zero energy state of the full problem 
could correspond to the uniqueness of an invariant wavefunction at 
$\infty$, (supersymmetric with respect to a free supercharge) adds to 
the hope to eventually find a simple reason for the (generally 
assumed) existence of the zero energy state.}, not exclude the 
existence of other non--zero, invariant, wavefunctions, annihilated by 
all ${\Q}_{\alpha}$).

What about the exact form of the full (non--asymptotic) 
wavefunction? If one wants to represent the fermionic degrees of 
freedom by creation and annihilation operators, e.g. by defining
{\def\theequation{\arabic{equation}}
\setcounter{equation}{13}
\begin{eqnarray}
  \lambda_{\alpha A} & := & \frac{1}{\sqrt{2}}\ (\Theta_{\alpha A} + 
  i \Theta_{\alpha + \frac{1}{2}\ s_{d},A}) \nonumber \\[0.5em]
  \frac{\partial}{\partial \lambda_{\alpha A}} & = & \frac{1}{\sqrt{2}}\ 
  (\Theta_{\alpha A} + 
  i \Theta_{\alpha + \frac{1}{2}\ s_{d},A}) \hspace{1ex} = 
  \hspace{1ex} \lambda_{\alpha A}^{\dagger}
\end{eqnarray}}
(where now $\alpha$ takes only half a many values as before!) explicit 
$Spin (d)$ invariance is lost, respectively nonlinearly realized [5]; 
some of the generators $J_{st}$ will not resprect the degree of 
homogeneity (mix the various components) of
{\def\theequation{15}
\begin{equation}
    \Psi = \psi + \psi_{\alpha A} \lambda_{\alpha A} + \frac{1}{2}\ 
    \psi_{\alpha A \beta B} \lambda_{\alpha A} \lambda_{\beta B} + 
    {\dots} + \frac{1}{\Lambda !}\ \psi_{\alpha_{1}A_{1},\dots , 
    \alpha_{\Lambda}A_{\Lambda}} \lambda_{\alpha_{1}A_{2}}\dots 
    \lambda_{\alpha_{\Lambda}A_{\Lambda}}\ .
\end{equation}} 
On the other hand, taking linear combinations of the Herimitan 
supercharges to form half as many nilpotent ones (on ${\mathcal{H}}$), 
and their Hermitian conjugates, the resulting supersymmetry algebra 
[5]
{\def\theequation{\arabic{equation}}
\setcounter{equation}{15}
\begin{eqnarray}
    \{Q_{\alpha}, Q_{\beta}^{\dagger}\} & \approx & \delta_{\alpha 
    \beta} H \nonumber \\
    \{Q_{\alpha}, Q_{\beta}\} & \approx & 0 \hspace{1ex} \approx 
    \hspace{1ex} \{Q_{\alpha}^{\dagger}, Q_{\beta}^{\dagger}\}
\end{eqnarray}}
will imply very simple properties,
{\def\theequation{17}
\begin{equation}
  (M \cdot \lambda)^2 = 0\ , \hspace{3ex}   (D \cdot \lambda)^2 = 
  0\ , \hspace{3ex} \{M \cdot \lambda, D \cdot \partial_{\lambda}\} 
  \approx 0
\end{equation}} 
when writing [33]
{\setcounter{equation}{17}
\begin{eqnarray}
Q_{\beta} & = & M_{\alpha\beta}^{(\beta)} \lambda_{\alpha A} + 
D_{\alpha A}^{(\beta)}\ \frac{\partial}{\partial \lambda_{\alpha A}} 
\hspace{1.5ex} = \hspace{1ex} M_{a} \lambda_{a} + D_{a} 
\partial_{\lambda_{a}} \hspace{1ex} = \hspace{1ex} M \cdot \lambda + 
D \cdot \partial_{\lambda} \nonumber \\
Q_{\beta}^{\dagger} & = & M_{\alpha\beta}^{(\beta)\dagger}
\partial_{\lambda_{\alpha A}}
+ 
D_{\alpha A}^{(\beta)\dagger}\lambda_{\alpha A} \hspace{1ex} = 
\hspace{1ex} M^\dagger \partial_{\lambda} + D^\dagger \lambda\ .
\end{eqnarray}}
The differential equations obtained componentwise from $Q_{\beta} 
\Psi = 0,$ \linebreak $Q_{\beta}^\dagger \Psi = 0$,
\setcounter{equation}{18}
\begin{eqnarray}
  D_{a_{2k}} \psi_{a_{1},\dots , a_{2k}} & = & (2k - 1)\ 
  M_{[a_{1}}\psi_{a_{2},\dots , a_{2k-1}]}   \\
  M_{a_{2k}}^\dagger \psi_{a_{1},\dots , a_{2k}} & = & (2k-1)\ 
 D_{[a_{1}}\psi_{a_{2},\dots , a_{2k-1}]}^\dagger\ ,
\end{eqnarray}
were first (recursively) solved [33] in the form
\setcounter{equation}{20}
\begin{eqnarray}
    \psi_{2k} & = & \chi_{2k}^{[h]} + \chi_{2k}^{[in]} \\[0.5em]
    \psi_{2k} & = & \psi_{2k}^{(h)} + \psi_{2k}^{(in)}
\end{eqnarray}
where `$h$' indicates the general solution of the corresponding 
homogenous equation, and `$in$' a particular solution of the 
inhomogeneous equation (given in [33]).

This procedure of solving $Q \Psi = 0$ or $Q^\dagger \Psi = 0$ can be 
written more elegantly by defining [34]
{\def\theequation{23}
\begin{equation}
   A := I^\dagger \cdot \lambda \hspace{2ex} D^\dagger \cdot \lambda\ , 
   \hspace{4ex} B := I \cdot \partial_{\lambda} \hspace{2ex} D \cdot 
   \partial_{\lambda}
\end{equation}}
where $I_{\alpha A}$ is chosen such that
{\def\theequation{24}
\begin{equation}
  \{ M \cdot \lambda, I \cdot \partial_{\lambda}\} = - \unity\ .  
\end{equation}} 
Formally, all solutions of $Q^\dagger \Psi = 0$, respectively $Q\chi = 
0$, are of the form
{\def\theequation{25}
\begin{equation}
 \Psi = (\unity - A)^{-1} \Psi^{(h)}\ , 
 \hspace{3ex}\text{respectively} \hspace{1ex} \chi = (\unity - 
 B)^{-1} \chi^{[h]}
\end{equation}}
with
{\def\theequation{26}
\begin{equation}
   (M^\dagger \partial_{\lambda}) \Psi^{(h)} = 0\ , \hspace{3ex} (M 
   \cdot  \lambda) \chi^{[h]} = 0
\end{equation}}
(as if $Q^\dagger \Psi = 0$, let $\Psi^{(h)} := \Psi - A \Psi$ and 
verify that $M^\dagger \partial_{\lambda} \Psi + D^\dagger \lambda 
\Psi + I^\dagger \lambda M^\dagger \partial_{\lambda} D^\dagger 
\lambda \Psi = 0$), while it is also not difficult to show that 
each $\Psi$ of the form (27) does solve $Q^\dagger \Psi = 0$, 
respectively $Q\chi = 0$:
\setcounter{equation}{26}
\begin{eqnarray}
  Q (\unity - B)^{-1}  \chi^{[h]} & = & (\unity - B)^{-1} 
  Q\chi^{[h]} \nonumber \\
  & & +\ (\unity - B)^{-1} [M \cdot \lambda + D \cdot 
  \partial_{\lambda}, I \cdot \partial_{\lambda} D \cdot \nonumber \\
  & & \cdot\
  \partial_{\lambda}\ (\unity - B)^{-1} Q\chi^{(h)} \\[0.5em]
  & = & (\unity - B)^{-1} (D \partial_{\lambda} (\unity - B)  \nonumber\\
  & & +\ \{M 
  \cdot \lambda, I \cdot \partial_{\lambda}\}\ D \cdot 
  \partial_{\lambda} I \cdot \partial_{\lambda} \{M \cdot \lambda, D 
  \cdot \partial_{\lambda}\} \nonumber \\
  & & +\ D \cdot \partial_{\lambda} B)\, (\unity - B)^{-1} \chi^{[h]} 
  \approx 0\ .
\end{eqnarray} 

\newpage

\subsection*{Excercise 5:} (cp. [40], [41])

\smallskip

Calculate $\{ Q_{\beta} , Q_{\beta'} \}$ when adding
$$
m_3 \sum_{i=1}^3 q_{iA} (\gamma^{123} \, \gamma^i)_{\beta\alpha} \, 
{\Theta}_{\alpha A} + m_6 \sum_{\mu = 4}^9 q_{\mu A} (\gamma^{123} \, 
\gamma^{\mu})_{\beta\alpha} \, {\Theta}_{\alpha A}
$$
to the r.h.s. of $Q_{\beta}$ in (3).

In particular, verify that the anticommutator of the above terms with the 
derivative part in the supercharges yields
\begin{eqnarray}
&&m_3 (-i (q_{iA} \, \nabla_{jA'} - q_{jA} \, \nabla_{iA}))(\gamma^{123} \, 
\gamma^{ij})_{\beta\beta'} \nonumber \\
&+ &m_6 (-i (q_{\mu A} \, \nabla_{\nu A} - q_{\nu A} \, \nabla_{\mu A})) 
(\gamma^{123} \, \gamma^{\mu\nu})_{\beta\beta'} \nonumber \\
&+ &2i \, \frac{\Theta_{\alpha A}}{32} \, \Theta_{\alpha' A} \{ (-6m_3 + 
12m_6) \, \delta_{\beta\beta'} \, \gamma_{\alpha\alpha'}^{123} \nonumber \\
&+ &(-m_3 + 6m_6) (\gamma^{123} \, \gamma^{ij})_{\beta\beta'} \, 
\gamma_{\alpha\alpha'}^{ij} \nonumber \\
&+ &(3m_3 + 2m_6) (\gamma^{123} \, \gamma^{\mu\nu})_{\beta\beta'} \, 
\gamma_{\alpha\alpha'}^{\mu\nu} \} \nonumber
\end{eqnarray}
(use that the $(\beta \leftrightarrow \beta')$ symmetric, $[\alpha 
\leftrightarrow \alpha']$ antisymmetric, part of $\gamma_{\beta'\alpha'}^i 
(\gamma^{123} \, \gamma^i)_{\beta \alpha}$, resp. 
$\gamma_{\beta'\alpha'}^{\mu} (\gamma^{123} \, \gamma^{\mu})_{\beta\alpha}$, 
can only contain terms proportional to the ones written above inside the curly 
bracket); recalling that the coefficient of $\Theta \gamma^{\cdot\cdot} \Theta$ 
in the 
generators of Spin (3), resp. Spin (6), should be $\frac{1}{4}$, relative to 
$q_{\cdot} \, \nabla_{\cdot} - q_{\cdot} \, \nabla_{\cdot}$ (cp. (3)), deduce 
that if (and only if) $m_6$ is equal to $-\frac{1}{2} \, m_3$, the above 
expression, apart from contributing to the Hamiltonian (a term $- 
\frac{3m_3}{4} \, i \, \Theta \, \gamma^{123} \, \Theta$), can be written 
solely in terms of symmetry-generators
$$
\left( m_3 \, J_{ij} (\gamma^{123} \, \gamma^{ij})_{\beta\beta'} - 
\frac{m_3}{2} \, J_{\mu\nu} (\gamma^{123} \, \gamma^{\mu\nu})_{\beta\beta'} 
\right) \, .
$$
The derivative-free extra contributions to (4) are all proportional to 
$\delta_{\beta\beta'}$, adding to the potential in (1)
\begin{eqnarray}
&&m_3^2 \, q_{iA} \, q_{iA} + \frac{m_3^2}{4} \, q_{\mu A} \, q_{\mu A} + m_3 
\, f_{ABC} \, \epsilon_{ijk} \, q_{iA} \, q_{jB} \, q_{kC} \nonumber \\
&= &Tr \left\{ \frac{m_3^2}{4} \, X^{\mu} \, X^{\mu} + (m_3 \, X_j - i \, 
\epsilon_{jk\ell} \, X_k \, X_{\ell})^2 \right\} \nonumber \\
&+ &\frac{1}{2} \, Tr \, [X_j , X_k] [X_j , X_k] \, ; \nonumber
\end{eqnarray}
here, $X_{\cdot} = i q_{\cdot A} \, T_A$.

\bigskip

\subsection*{Excercise 6:} (cp. [34], [42])

\smallskip

Let $\lambda_a$ and $\lambda_a^{\dagger} = \frac{\partial}{\partial 
\lambda_a}$ ($a = 1 , \ldots , \Lambda$) be fermionic creation, resp. 
annihilation-operators; i.e.  $\{ \lambda_a , \lambda_b \} = 0 = \left\{ 
\frac{\partial}{\partial \lambda_a} , \frac{\partial}{\partial \lambda_b} 
\right\}$, $\left\{ \lambda_a , \frac{\partial}{\partial \lambda_b} \right\} = 
\delta_{ab}$, $\frac{\partial}{\partial \lambda_a} \mid 0 \rangle = 0$ (for 
all $a$, and $b$).

\smallskip

Define [34]
$$
* := \sum_{m=0}^{\Lambda} (-)^m \frac{\epsilon_{a_1 \ldots a_{\Lambda}}}{m! 
(\Lambda - m)!} \, \lambda_{a_1} \cdot \ldots \cdot \lambda_{a_m} \, 
\partial_{\lambda_{a_{\Lambda}}} \ldots \partial_{\lambda_{a_{m+1}}} \, .
$$
Show that on the $p$-fermion-sector, i.e. on states $\vert \Psi_p \rangle = 
\Psi_{a_1 \ldots a_p} \, \lambda_{a_1} \cdot \ldots \cdot \lambda_{a_p} \, 
\vert 0 \rangle$, one has
$$
\partial_{\lambda_a} * \vert \Psi_p \rangle = (-)^{\Lambda - p} * \lambda_a 
\vert \Psi_p \rangle
$$
$$
\lambda_a * \vert \Psi_p \rangle = (-)^{\Lambda - p + 1} * 
\partial_{\lambda_a} \vert \Psi_p \rangle \, ,
$$
as well as $*^2 \, \vert \Psi_p \rangle = (-)^{p(\Lambda - p) + \Lambda} \vert 
\Psi_p \rangle$; thus, (note the sign-errors in equations (17)-(19) of [34])
$$
H_{ab} \, \lambda_a \, \frac{\partial}{\partial \lambda_b} \, * = * H_{ab} \, 
\lambda_a \, \frac{\partial}{\partial \lambda_b}
$$
$$
(F_{ab} \, \lambda_a \, \lambda_b + G_{ab} \, \partial_{\lambda_a} \, 
\partial_{\lambda_b}) \, * = - * (F_{ab} \, \lambda_a \, \lambda_b + G_{ab} \, 
\partial_{\lambda_a} \, \partial_{\lambda_b}) \, .
$$
For the $SU(N)$-invariant matrix-models (1)
$$
F_{\alpha A , \beta B} = \delta_{\alpha\beta} \, f_{ABE} (q_{d-1,E} + i \, 
q_{d,E})
$$
$$
G_{\alpha A , \beta B} = - F_{\alpha A , \beta B}^*
$$
$$
H_{\alpha A , \beta B} = H_{\alpha A , \beta B} (q_{jE})_{j=1,\ldots ,d-2} \, 
.
$$
Therefore,  if one defines $\widehat{q}_{dE} := -q_{dE}$, $\widehat{q}_{s\ne d 
, E} = q_{sE}$, $H$ will commute with the joint action of $*$ and $\widehat{ \ 
} \, $, 
hence $H ( * \Psi (\widehat q)) = 0$ if $H \, \Psi (q) = 0$. For $d=2$ and 
even $N$ this means (cp. [42]; there, $*$ is taken to be the Hodge-operator -- 
which corresponds to taking $(-)^{m(\Lambda - m)}$, instead of $(-)^m$, in the 
above definition of $*$) that the corresponding index (of $H$) trivially 
vanishes, as for odd $\Lambda \quad$ $* \Psi (\widehat q)$ will be fermionic if 
$\Psi (q)$ is bosonic (and vice versa).

\newpage
  
  \begin{center}
  {\bf IV}
\end{center}

Consider 3 traceless, antihermitean $N \times N$ matrices 
$X_a(t),\ t \in 
(-\infty, + \infty)$, developping in time according to the equations
\def\theequation{1}
\begin{equation}
 \dot{X_a} = \epsilon_{abc} X_bX_c - m X_a \, 
.~~\label{eq:1}\end{equation}
The stationary points of this flow are representations of $su(2)$, i.e. 
$X_a = mJ_a$,
\def\theequation{2}
\begin{equation}[J_a,\, J_b] = \epsilon_{abc}J_c\, .\end{equation}
The question is: given 2 such representations, $\rho_+$ and $\rho_-$,
under 
which circumstances do there exist solutions $X_a(t)$ of (1) approaching
the 
representation $\rho_+$ as $t \to + \infty$ and (being conjugate to)
$\rho_-$ as 
$t \to -\infty\, $?

Denoting the space of such solutions by ${\mathcal M}(\rho_{-},\,
\rho_{+})$, 
Kronheimer 
[10], in parts building on work of Slodowy [9], proved that
\def\theequation{3}
\begin{equation}{\mathcal M}(\rho_-,\, \rho_+) = {\mathcal N} (\rho_-)
\cap S(\rho_+) \, 
,\end{equation}
where the r.h.s. is well known from singularity theory related to Lie
algebras 
[9]. In the main part of this lecture, based on joint work with C.
Bachas and B. 
Pioline (see [11]; in particular concerning the physical relevance of
(1), 
(3)) I will discuss (3):

Take 
\def\theequation{4}
\begin{equation}
h =\left(
\begin{array}{cc}
1 & 0 \\
0 & -1 \\
\end{array}
\right),
\ 
x =\left(
\begin{array}{cc}
0 & 1 \\
0 & 0 \\
\end{array}
\right),
\  
y = \left(
\begin{array}{cc}
0 & 0 \\
1 & 0 \\
\end{array}
\right)
\end{equation}
as generators of $s\ell(2,\C)$, the complexification of $su(2)$; denote
by 
$H_\pm 
:= \rho_\pm(h)$, $X_\pm := \rho_\pm(x)$,\ $Y_\pm:=\rho_\pm(y)$, the 
corresponding 
$N
\times N$ matrices in the representation $\rho_\pm$, i.e. satisfying the 
same 
commutation relations as those following from (4),
\def\theequation{5}
\begin{equation}
[x,y]=h,\ [h,x]=2x,\ [h,y]=-2y\, .
\end{equation}
$ {\mathcal N}(\rho_\pm)$ is then defined as the orbit of $Y_\pm$ under
the 
complexified 
gauge group, $SU(N)_{\C}   =  SL(N,\C)$:
\def\theequation{6}
\begin{equation}
{\mathcal N}(\rho_{- \atop (+)}):  =  \{g Y_{- \atop (+)} g \mid g \in
SL(N,\C) 
\}
\end{equation}
while
\def\theequation{7}
\begin{equation}
S (X_{+ \atop{(-)}}) = Y_{+ \atop(-)} + Z (X_{+ \atop{(-)}})
\end{equation}
where
\def\theequation{8}
\begin{equation}
Z(X_{+ \atop{(-)}}) : = \{A \in s\ell(N,\C) \mid [A, X_{+ \atop{(-)}}]=0\}
\end{equation}
is the centralizer of $X_{+ \atop {(-)}}$.

\medskip\noindent Example $(N=3)$:

Let $\rho_-$ be the irreducible 3-dimensional representation of $su(2)$,
and 
$\rho_+=2 \oplus 1$ the direct sum of the irreducible 2-dimensional one,
and 
the 
trivial 1-dimensional (putting all $J_a=0$). Then one has
%\begin{equation} 
\begin{eqnarray*}
Y_- = \sqrt{2} \left(
\begin{array}{ccc}
0 & 0 & 0 \\
1 & 0 & 0 \\
0 & 1 & 0 \\
\end{array}
\right) 
&
Y_+  =\left(
\begin{array}{cccc}
0 & 0  & \vline & 0 \\
1 & 0  & \vline & 0 \\
\hline
0 & 0  & \vline & 0 \\
\end{array}
\right)
\end{eqnarray*}
\begin{eqnarray*}
X_- = \sqrt{2} \left(
\begin{array}{ccc}
0 & 1 & 0 \\
0 & 0 & 1 \\
0 & 0 & 0 \\
\end{array}
\right) 
%\qquad
&
X_+  =\left(
\begin{array}{cccc}
0 & 1  & \vline &  \\
0 & 0  & \vline &  \\
\hline
  &    & \vline & 0 \\
\end{array}
\right)\\ 
\end{eqnarray*}
{\def\theequation{\arabic{equation}}
\setcounter{equation}{8}
\begin{eqnarray}
H_- = \left( 
\begin{array}{ccc}
2 & 0 & 0 \\
0 & 0 & 0 \\
0 & 0 & -2 \\
\end{array}
\right) 
&
H_+  =\left(
\begin{array}{cccc}
1 & 0  & \vline & \\
0 & -1 & \vline &  \\
\hline
  &    & \vline & 0 \\
\end{array}
\right)  \, .
\end{eqnarray}}
%\end{equation}
In this example,
\def\theequation{10}
\begin{equation}
S(\rho_+) = \left \{s =
\left(
\begin{array}{ccc}
a & b & c \\
1 & a  & 0\\
0 & e & -2a \\
\end{array}
\right)
\mid a,b,e,c \in \C \right \}\, ,
\end{equation}
as can be found either by a simple explicit computation, or using the
general 
fact that $s\ell(N,\C)$ decomposes, under the adjoint action of $\rho_{+ 
\atop{(-)}}$ into irreducible representation spaces of $s\ell(2,\C)$,
each of 
which 
contains exactly one 1-parameter family of elements of $Z(X_{+
\atop{(-)}})$; 
in the above case, $s\ell(3,\C)$ decomposes, under the action of
$(Y_+=E_{21},\ 
X_+ 
= E_{12}, \, H_+ = E_{11} - E_{22})$ into one 3-dimensional
representation 
space 
($\rho_+$ itself, contributing $\C \cdot X_+$ to $Z(X_+)$), two 
2-dimensional\linebreak ones: 
spanned by $E_{23}$ and $[E_{12},  E_{23}]=E_{13} \in Z(E_{12})$, resp. 
$E_{31}$ and $[E_{12}, E_{31}]= -E_{32} \in Z (E_{12})$, and one
1-dimensional 
one $(\C \cdot (E_{11} + E_{22} - 2 E_{33})\in Z(E_{12}))$. Instead of 
computing 
${\mathcal N} (\rho_-)$ explicitly, ${\mathcal N}(\rho_-) \cap S
(\rho_+)$ can, in the 
above example, be determined by simply demanding $s^3=0,\ s^2 \not = 0$
for the 
elements in (10); this gives $b=-3a^2,\ ec=8a^3$, i.e.
{\def\theequation{\arabic{equation}}
\setcounter{equation}{10}
\begin{eqnarray}
\lefteqn{{\mathcal N} (\rho_-) \cap S (\rho_+) =}\\ 
 =& \left \{ \left (\begin{array}{ccc}
a & -3a^2 & c \\
1 & a  & 0\\
0 & e & -2a \\
\end{array}
\right )
 \vrule {
 a,c,e \in \C \atop{ec=8a^3}} \right\} \ .\nonumber
 \end{eqnarray}}
 According to (3), ${\mathcal{M}}(\rho_-,\rho_+)$ is therefore the
4-dimensional 
(singular) space (11). Let me now sketch (part of) the proof of (3) (cp
[10]): 
One first 
`gauges' (1) by introducing a 4-th traceless, antihermitean, $N \times 
N$ 
matrix, $X_0$, and going over to the equations 
\def\theequation{12}
 \begin{equation}
 \dot{X_a} + [X_0,X_a] = \frac{1}{2}\,  \epsilon_{abc} [X_{b}, X_{c}] -
m X_a\, 
.
 \end{equation} 
Due to their invariance under 
\def\theequation{13}
\begin{equation} 
\begin{array}{rcl}
 X_a \to \tilde X_a & = & U(t)X_aU^{-1}(t)\, ,\\
X_0 \to \tilde X_0 & = & U X_0 U^{-1} - \dot U U^{-1}\, ,
\end{array}
\end{equation} 
a solution $\tilde X_a$ of (1) may be obtained from a solution $X_a$ of
(12) by 
choosing $U$ in (13) such that $\tilde X_0 = 0$. (12) is then split into
one 
complex equation (from now on, $m=2$)
\def\theequation{14}
\begin{equation}
\dot \beta + 2 \beta + 2 [\alpha,\beta]=0 \, ,
\end{equation}
and one real equation, 
\def\theequation{15}
\begin{equation}
\frac{d}{dt} \, (\alpha + \alpha^\dagger) + 2 (\alpha + \alpha^\dagger)
+ 2 
[\alpha, 
\alpha^\dagger] 
+ 2[\beta,\beta^\dagger] = 0 \, .
\end{equation} 
 Due to $\alpha := \frac{1}{2} (X_0 - i X_3)$ and $\beta := -
\frac{1}{2} (X_1 
+ i X_2)$ no longer having to obey any (anti)hermiticity conditions, the 
gauge-invariance of (14) is enhanced to complex (!) gauge
transformations 
%\end{document}
\def\theequation{16}
\begin{equation} 
 \begin{array}{rcl}
 \alpha &  \to & g \alpha g^{-1} - \frac{1}{2} \, \dot g g^{-1}\\
 & & \qquad \qquad \qquad \qquad g \in SL(N,\C)\\
 \beta  & \to  & g \beta g^{-1}\, .
 \end{array}
 \end{equation}
 Kronheimer [1] then proved that any solution of (14) (with the required 
boundary conditions) is gauge equivalent to
{\def\theequation{\arabic{equation}}
\setcounter{equation}{16}
 \begin{eqnarray}
 \alpha_-(t) & = & \frac{1}{2}\,  H_-,\ \beta_- (t) = Y_- \quad
\mbox{\rm for}\ 
t 
\in (-\infty,0] \nonumber \\
 \alpha_+ (t) & = & \frac{1}{2}\,  H_+,\ \beta_+ (t) = Y_+ + e^{-2t}
e^{-t \, 
ad\, 
H_+}Z_+ \nonumber\\ 
&&\qquad\mbox{\rm for}\  t \in [0,+ \infty)\, ,
 \end{eqnarray}}
with $Z_+ \in Z(X_+)$. Stated the other way round (actually 0 may be
replaced 
by any finite time, in (17)): for any given solution $(\alpha, \beta)$
of (14) 
there exist $g_+$ and $g_-$ (approaching the identity, resp. a constant
group  
element, at $t=+ \infty$, resp. $t = - \infty$) 
such that, for any finite $t$,
{\def\theequation{\arabic{equation}}
\setcounter{equation}{17}
 \begin{eqnarray}
 \beta    & = & g_+^{-1} ( Y_+ + e^{-(2 + \, ad \, H_+)t} Z_+)g_+ 
\nonumber\\
 2\alpha  & = & g_+^{-1} H_+ g_+ + g_+^{-1} \dot{g}_+
 \end{eqnarray}}
 AND
 {\def\theequation{\arabic{equation}}
\setcounter{equation}{18}
 \begin{eqnarray*}
 \beta    & = & g_-^{-1} Y_- g_- \\
 2 \alpha & = & g_-^{-1} H_- g_- + g_-^{-1} \dot{g}_-\, .
 \end{eqnarray*}}
 This means that for any finite $t$, 
\def\theequation{19}
 \begin{equation}
 Y_+ + e^{-(2+\, ad\, H_+)t}Z_+ \, , 
 \end{equation}
 which is $\in  S(\rho_+)$, must be gauge-equivalent to $Y_-$, i.e. must
be 
$\in {\mathcal{N}} (\rho -)$. Putting $t=0$, and assuming (proved in
[10]) 
that the gauge--invariance can always be used to satisfy (15), this 
gives the desired correspondance between solutions of (1), 
interpolating between $\rho_{-}$ and $\rho_{+}$, and $(Y_{+} + Z(X_{+})) 
\cap {\mathcal{N}}(Y_{-})$. Letting $t \to + \infty$, 
while noting that $(2+ \,ad\,  
H_+)$ is strictly 
positive
{\renewcommand{\thefootnote}{\fnsymbol{footnote}}\footnote[1]{in the 
previous 
example one would have
 \begin{eqnarray*}
[H_+ = E_{11} - E_{22},
 \begin{array}{c}
 X_+\\
 E_{13}\\
 E_{32}\\
 E_{11} + E_{22} - 2E_{33}
 \end{array}  ]= 
 \begin{array}{c}
 2X_+\\
 1 \cdot E_{13}\\
 1 \cdot E_{32}\\
 0
 \end{array}
 \end{eqnarray*}
}
 on $Z(X_+)$, one finds that $Y_+$ (hence ${\mathcal{N}}(\rho_+)$!) must
actually 
be contained in the closure of ${\mathcal{N}} (\rho_ -)$ (for 
${\mathcal{M}} (\rho_ -, 
\rho_+)$ to be non-empty). If this condition is fulfilled, the dimension
of 
${\mathcal{M}}$, due to $S_+$ and ${\mathcal{N}}_-$ meeting 
transversely, can be computed as follows:
{\def\theequation{\arabic{equation}}
\setcounter{equation}{19}
 \begin{eqnarray}
\lefteqn{ \dim (S_+ \cap {\mathcal{N}}_-)} \\
& = & \dim S_+ + \dim {\mathcal{N}}_- - \dim (S_+ \cap 
{\mathcal{N}}_-)\nonumber\\
 & = & \dim S_+ - \dim S_- \, .\nonumber
 \end{eqnarray}
 As a general $N \times N \hspace{2ex} SU(2)$ representation is a 
 direct sum of irreducible ones,
  \begin{equation}
\rho = \sum_{j\in {\N}/2} n_{j} [j]
  \end{equation}
  with $\dim [j] = 2j + 1 \in {\N}, \rho_{\pm}$ correspond to 
  partitions $P_{\pm}$ of $N$, hence are associated to 
  Young--tableaux(`s) $T_{\pm}$, the number of boxes in row $k$ 
  corresponding to the dimension of the $k$'s representation in 
  (21), --- having ordered them with the largest $[j]$ first. The 
  condition that ${\mathcal{N}}(g_{+}) \subset 
  \overline{{\mathcal{N}}(g_{-})}$ then translates (see e.g. [45, 11])
into 
  the condition that, for each $p = 1, 2, \ldots$, the number of boxes 
  contained in the first $p$ columns must not decrease (when going 
  from $T_{-}$ to $T_{+}$). Another way to state the same condition 
  is to compare the null--spaces of powers of $Y_{-}$ to those of 
  $Y_{+}$ (which, for each given power, must not decrease).
  
  \newpage 
  
\subsection*{Excercise 7:}

\smallskip

In [44] it has been shown that if $X_j = \partial_1 \, f_j \, \partial_2 - 
\partial_2 \, f_j \, \partial_1$ ($j=1 , \ldots , 5$) are 5 divergence-free 
vectorfields, their totally antisymmetrized product $s_5 (X_1 , \ldots , X_5)$ 
will, in contrast with all other $M$-commutators $(M>2)$, again be a (generally 
non-zero) divergence-free vector-field, and 
$$
ad \, s_5 (X_1 , \ldots , X_5) = s_5 (ad \, X_1 , \ldots , ad \, X_5) \, ,
$$
as well as
$$
[X_6 , s_5 (X_1 , \ldots , X_5)] = \sum_{j=1}^5 s_5 (X_1 , \ldots , X_{j-1} , 
[X 
, X_j] , X_{j+1} , \ldots , X_5) \, ;
$$
correspondingly one may define
$$
\{ f_1 , \ldots , f_5 \}_5 := \left\vert \begin{matrix}
\partial_1 f_1 &\partial_1 f_2 &\partial_1 f_3 &\partial_1 f_4 &\partial_1 f_5 
\\
\partial_2 f_1 &\ldots &\ldots &\ldots &\partial_2 f_5 \\
\partial_1^2 f_1 &\ldots &\ldots &\ldots &\partial_1^2 f_5 \\
\partial_{12}^2 f_1 &\ldots &\ldots &\ldots &\partial_{12}^2 f_5 \\
\partial_2^2 f_1 &\ldots &\ldots &\ldots &\partial_2^2 f_5 \\
\end{matrix} \right\vert
$$
for functions on $T^2$, with $\{ \ldots \}_5$ and the ordinary Poisson-bracket 
$\{ f,g \} := \partial_1 f \, \partial_2 g - \partial_2 f \, \partial_1 g$ 
satisfying various (mixed) identities, --  which will be quite relevant 
for the understanding of 2- and 5-branes in $M$-theory.

\medskip

{\bf Conjecture} (Approximation of 5-commutator of functions by 5-commutator 
of matrices):

\smallskip

Let $\gamma (\stackrel{\rightharpoonup}{m}_1 , \ldots , 
\stackrel{\rightharpoonup}{m}_5)$ be defined via
$$
\{ f_{\stackrel{\rightharpoonup}{m}_1} , \ldots , 
f_{\stackrel{\rightharpoonup}{m}_5} \}_5 = \gamma 
(\stackrel{\rightharpoonup}{m}_1 , \ldots , \stackrel{\rightharpoonup}{m}_5) \, 
e^{i \sum_{j=1}^{5} \stackrel{\rightharpoonup}{m}_j \cdot 
\stackrel{\rightharpoonup}{\varphi}} \, ,
$$
$f_{\stackrel{\rightharpoonup}{m}_j} = e^{i \stackrel{\rightharpoonup}{m}_j 
\stackrel{\rightharpoonup}{\varphi}}$, and $T_{\stackrel{\rightharpoonup}{m}} 
:= 
w^{\frac{1}{2} m_1 m_2} \, g^{m_1} \, h^{m_2}$ (with $w,g,h$ as in II (55); 
$M=1$ 
for simplicity); then, as $N \rightarrow \infty$,
$$
s_5 (T_{\stackrel{\rightharpoonup}{m}_1} , \ldots , 
T_{\stackrel{\rightharpoonup}{m}_5}) = C_N \, \gamma 
(\stackrel{\rightharpoonup}{m}_1 , \ldots , \stackrel{\rightharpoonup}{m}_5) \, 
T_{\sum_1^5 \stackrel{\rightharpoonup}{m}_j} \left( 1+0 \left( \frac{1}{N} 
\right)\right) \, .
$$

  \newpage
  
\subsection*{References}
\begin{enumerate}
\item[\mbox{[1]}] J.\ Hoppe; ``Quantum Theory of a Massless Relativistic 
Surface\ldots'' (1982): MIT Ph.D. Thesis. \\ J.\ Goldstone; 
unpublished.
\item[\mbox{[2]}]J.\ Hoppe; ``Quantum Theory of a Relativistic Surface''
in 
``Constraint's Theory and Relativistic Dynamics'', {\em World 
Scientific}
1987 (eds. G.\ Longhi, L.\ Lusanna).
\item[\mbox{[3]}] J.\ Hoppe; ``Elementary Particle Research Journal''
(Kyoto) 
{\bf 80}, 3 (1989).
\item[\mbox{[4]}] R.\ Flume; {\em Ann.Phys.} {\bf 164} (1985) 189. \\
M.\ 
Claudson, M.\ Halpern; {\em Nucl.Phys.} {\bf 250} (1985) 689. \\ M.\
Baake, P.\ 
Reinicke, V.\ Rittenberg; {\em J.Math.Phys.} {\bf 26} (1985) 1070.
\item[\mbox{[5]}] B.\ de\,Wit, J.\ Hoppe, H.\ Nicolai; {\em Nucl.Phys.}
{\bf 
B 305} (1988) 545.
\item[\mbox{[6]}] E.\ Witten; {\em Nucl.Phys.} {\bf B 460} (1996) 335. 
\\ T.\ Banks, 
W.\ Fischler, S.\ Shenker, L.\ Susskind; {\em Phys.Rev.} {\bf D 55} 
(1997) 5112.
\item[\mbox{[7]}] M.\ Bordemann, J.\ Hoppe; {\em Phys.Lett.} {\bf B 317}
(1993) 
315.
\item[\mbox{[8]}] B.\ de\,Wit, M.\ L\"uscher, H.\ Nicolai; {\bf B 320}
(1989) 135.
\item[\mbox{[9]}] P.\ Slodowy; {\em Lecture Notes in Mathematics} 
{\bf 815}, 
Springer 1980.
\item[\mbox{[10]}] P.\ Kronheimer; {\em J.Diff.Geom.} {\bf 32} (1990) 473.
\item[\mbox{[11]}] C.\ Bachas, J.\ Hoppe, B.\ Pioline; hep--th/007067.
\item[\mbox{[12]}] J.\ Hoppe; {\em Phys.Lett.} {\bf B 335} (1994) 41.
\item[\mbox{[13]}] M.\ Bordemann, J.\ Hoppe; {\em J.Math.Phys.} {\bf 39} 
(1998) 683.
\item[\mbox{[14]}] J.\ Hoppe; hep--th/9407103.
\item[\mbox{[15]}] J.\ Hoppe; hep--th/9503069.
\item[\mbox{[16]}] D.G.\ Currie, T.F.\ Jordan, E.C.G.\ Sudarshan; {\em Rev. 
Mod. Phys.} {\bf 35} (1965) 350; and H.\ Leutwyler; {\it Nuovo Cim.} {\bf 37} 
(1965) 556.
\item[\mbox{[17]}] J.\ Hoppe, T.\ Ratiu; {\em Class and Quant. Gravity}
{\bf 
14} (1997) L45.
\item[\mbox{[18]}] A.\ Sugamoto; {\em Nucl.Phys.} {\bf B 215} (1983) 381.
\item[\mbox{[19]}] M.\ Bordemann, J.\ Hoppe; {\em Phys.Lett.} {\bf 325}
(1994) 
359.
\item[\mbox{[20]}] T.R.\ Morris; From first to second quantized string theory; 
{\it Phys. Lett. B} {\bf 202:222} (1988); and D.L.\ Gee, T.R. Morris; From 
first to second quantized string theory. 3. Gauge fixing and quantization; {\it 
Nucl. Phys. B} {\bf 331:694} (1990).
\item[\mbox{[21]}] J.\ Hoppe; hep--th/9311059.
\item[\mbox{[22]}] J.\ Hoppe; MIT Ph.D. Thesis (see {[1]}). Since 1999
available 
\hbox{under} \hspace{1ex} 
http:$\slash\slash$www.aei-potsdam.mpg.de$\slash$$^\sim$hoppe
\item[\mbox{[23]}] J.\ Hoppe; {\em Helvetia Physica Acta} Vol. 70 (1997) 
hep/th
\item[\mbox{[24]}] J.\ Goldstone; unpublished notes (1985).
\item[\mbox{[25]}] D.\ Fairlie, P.\ Fletcher, C.\ Zachos; {\em Phys.Lett.} {\bf 
B 
218} (1989) 203.\\
J.\ Hoppe; {\em Phys.Lett.} {\bf B 215} (1988) 706.
\item[\mbox{[26]}] M.\ Bordemann,  J.\ Hoppe, P.\ Schaller, M.\ 
Schlichenmaier; \\ {\em Comm.Math.Phys.} {\bf 138} (1991) 209.
\item[\mbox{[27]}] M.\ Bordemann, E.\ Meinrenken, M.\ Schlichenmaier; 
CMP  {\bf 165} (1994) 281.
\item[\mbox{[28]}] J.\ Hoppe, S.--T.\ Yau; Matrix Harmonics on $S^2$, 
{\em Comm.Math.Phys.} {\bf 195} (1998) 67.
\item[\mbox{[29]}] A.\ Smilga; {\em Nucl. Phys.} {\bf B 266} (1986) 45. \\
A.\ Smilga; Super--Yang--Mills quantum mechanics and supermembrane 
spectrum (1990) {\em Proc. 1989 Trieste Conf.} ed M.\ Duff, C.\ Pope and
E. 
Sezgin (Singapore: World Scientific). 
\item[\mbox{[30]}] J.\ Fr\"ohlich and J.\ Hoppe; {\em Commun. 
Math. Phys.} {\bf 191} (1998) 613. \\
J.\ Hoppe and S.--T.\ Yau; Absence of zero energy states in the 
simplest $d=3\;(d=5?)$ matrix models, hep--th/9806152. \\
J.\ Fr\"ohlich, G.M.\ Graf, D. Hasler, J.\ Hoppe and S.--T.\ Yau; 
Asymptotic form of zero energy wave functions in supersymmetric 
matrix models, hep--th/9904182.
\item[\mbox{[31]}] M.\ Halpern, C.\ Schwartz; Int.J.Mod.Phys. {\bf A 
13} (1998) 4367.
\item[\mbox{[32]}] G.M.\ Graf and J.\ Hoppe, Asymptotic ground 
state for $10$--dimensional reduced supersymmetric $SU(2)$ Yang--Mills 
theory, hep--th/980580.
\item[\mbox{[33]}] J.\ Hoppe; On the construction of zero 
energy states in supersymmetric matrix models, I + II,
hep--th/9709132.
\item[\mbox{[34]}] J.\ Hoppe; On the construction of zero energy 
states in supersymmetric matrix models, III, 
hep--th/9711033.
\item[\mbox{[35]}] V.\ Kac, A.\ Smilga; {\em Nucl.Phys.} {\bf B 571} 
(2000) 515.
\item[\mbox{[36]}] M.\ Bordemann, J.\ Hoppe, R.\ Suter; Zero Energy 
States for $SU(N): \dots$, hep--th/9909191.
\item[\mbox{[37]}] J.\ Hoppe; Asymptotic Zero Energy States for $SU(N 
\ge 3)$, hep--th/9912163.
\item[\mbox{[38]}] J.\ Hoppe, J.\ Plefka; The Asymptotic Groundstate 
of $SU(3)$ Matrix Theory, hep--th/0002107.
\item[\mbox{[39]}] J.\ Hoppe, P.\ Schaller; Phys.Lett. {\bf B 237} 
(1990) 407.
\item[\mbox{[40]}] D.\ Berenstein, J.\ Maldacena, H.\ Nastase; Strings in flat 
space and $pp$ waves from ${\mathcal N} = 4$ Super Yang Mills; 
hep-th/0202021.
\item[\mbox{[41]}] K.\ Dasgupta, M.M.\ Sheikh-Jabbari, M.\ van Raamsdonk; 
Matrix Perturbation Theory For $M$-Theory On a $PP$-Wave; hep-th/\break  
0205185.
\item[\mbox{[42]}] G.M.\ Graf, D.\ Hasler, J.\ Hoppe; Vanishing index for 
supersymmetric 2-matrix model with odd dimensional gauge group; 
hep-th/0205285.
\item[\mbox{[43]}] J.\ Hoppe; Some Classical Solutions of Membrane Matrix 
Model Equations; Proceedings of the Carg\`ese Nato Advanced Study Institute, 
May 1997, Kluwer 1999.
\item[\mbox{[44]}] A.S. Dzhumadil'daev; $N$-commutators of vector fields; 
math. RA/ 0203036.
\item[\mbox{[45]}] H.\ Kraft, C.\ Procesi; Comm.Math.Helv. {\bf 57} 
(1982) 539.
\end{enumerate}

\end{document}